\documentclass[
  ,draft            
  ]
  {aipproc}
\layoutstyle{6x9}
\usepackage{psfig,epsfig,color}
\begin{document}
\title{Strange Hadron Resonances:\\
Freeze-Out Probes in Heavy-Ion Collisions
}

\author{C. Markert}{
  address={Yale University, New Haven, Connecticut 06520}
}

\author{G. Torrieri}{
  address={Department of Physics, University of Arizona, Tucson, AZ 85721}
}

\author{J. Rafelski}{
  address={Department of Physics, University of Arizona, Tucson, AZ 85721}
}

\begin{abstract}
Hyperon resonances are becoming an extremely useful tool allowing the study
of the properties of hadronic fireballs made in heavy ion collisions. Their 
yield, compared to  stable particles with the same quark composition, 
depends  on  hadronization conditions. The resonance's short lifetime 
makes them ideal probes of the fireball chemical freeze-out mechanisms.  An 
analysis of resonance abundance in heavy ion collisions should be capable 
of distinguishing between possible hadronization scenarios, in particular 
between sudden and gradual hadronization. In this paper, we review the 
existing SPS and RHIC experimental data on resonance production in heavy 
ion collisions, and discuss in terms of both thermal and microscopic models 
the yields  of the two observed resonances, $K^*$ and $\Lambda(1520)$. 
We show how freeze-out properties, namely chemical freeze-out temperature 
and the lifetime of the interacting hadron phase which follows, can be 
related  to resonance yields. Finally, we apply these methods to SPS and 
RHIC measurements, discuss the significance and interpretations of our 
findings, and suggest further measurements which may help
in clarifying existing ambiguities.
\end{abstract}
\maketitle
\section{Introduction and motivation}

\begin{figure}[tb]
\includegraphics[width=012cm]{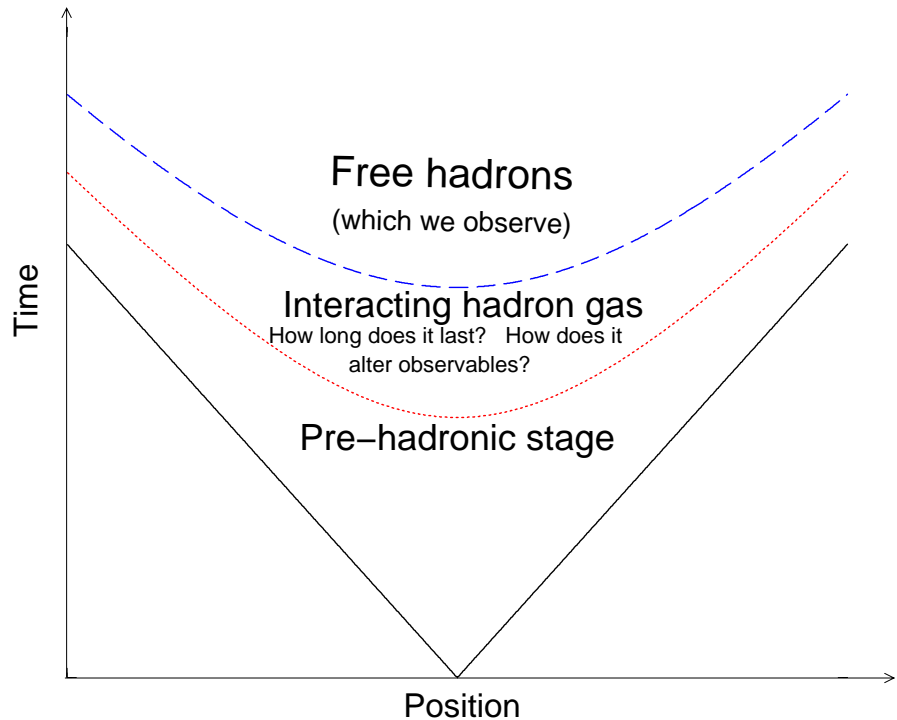}
\caption{ Stages of the space-time evolution of a
heavy ion collision.  At a certain moment in proper time (known as chemical freeze-out), hadrons emerge.
The system then evolves as an interacting hadron gas, until thermal
freeze-out, the point at which all elastic interactions cease as well. \label{problem}}
\end{figure}

The experimental measurement of short-lived hadron resonances can potentially
be very useful in clarifying some of the least understood aspects of heavy
ion collisions.
In general, evolution of a hot hadronic system proceeds according
to Fig. \ref{problem}.
When mesons and baryons emerge from a pre-hadronic state, presumably
quark gluon plasma, their abundances are
expected to be fixed by hadronization temperature and chemical fugacities.
This stage of fireball evolution is commonly
known as chemical freeze-out.  
After initial hadronization, the system may evolve as an interacting
hadron gas.
At a certain point (which can vary according to particle
species), thermal freeze-out, where hadrons stop interacting, is reached.

\begin{figure}[h]
\psfig{width=7.2cm,clip=1,figure=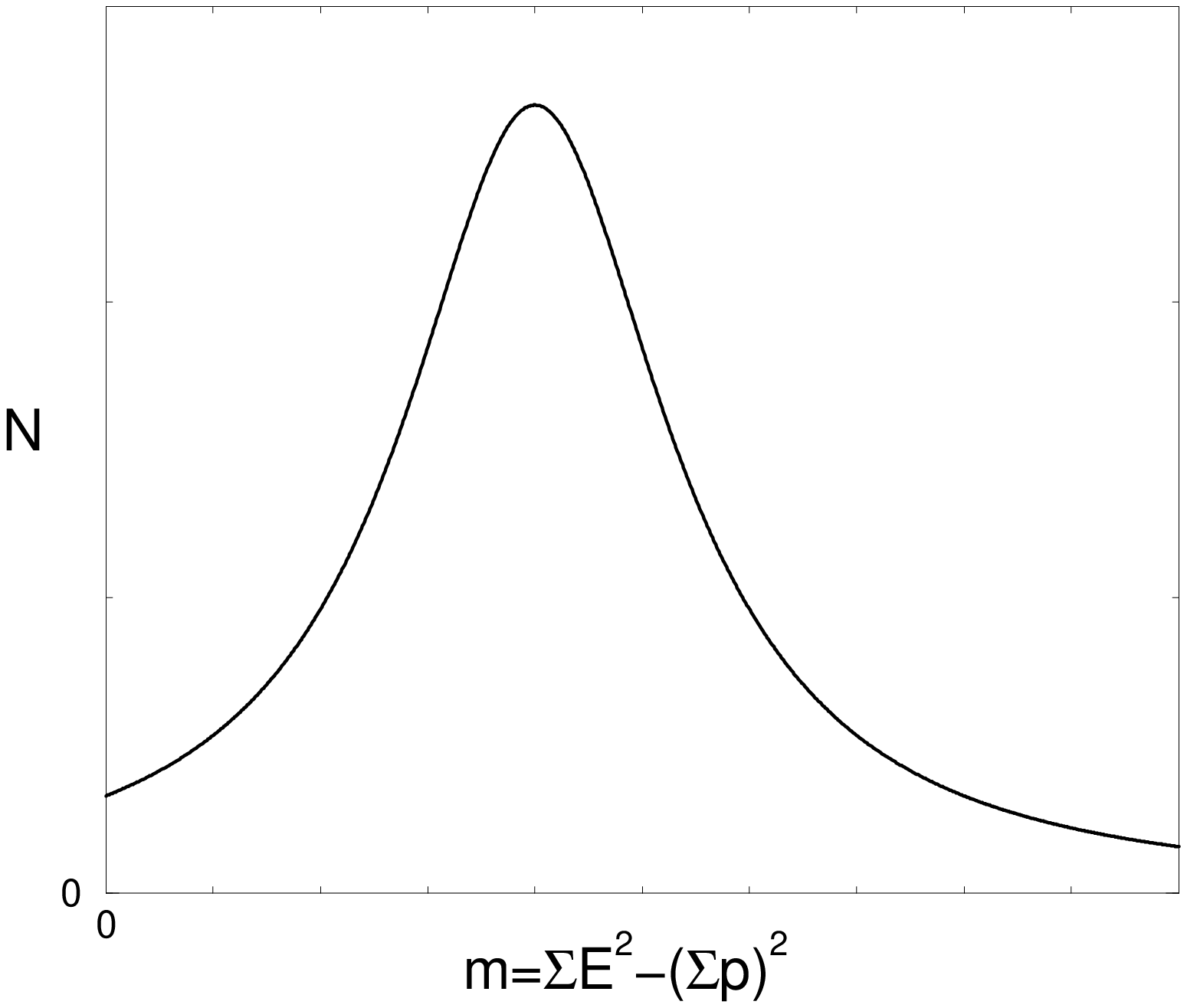}\hspace*{0.4cm}
\psfig{width=7.cm,clip=1,figure=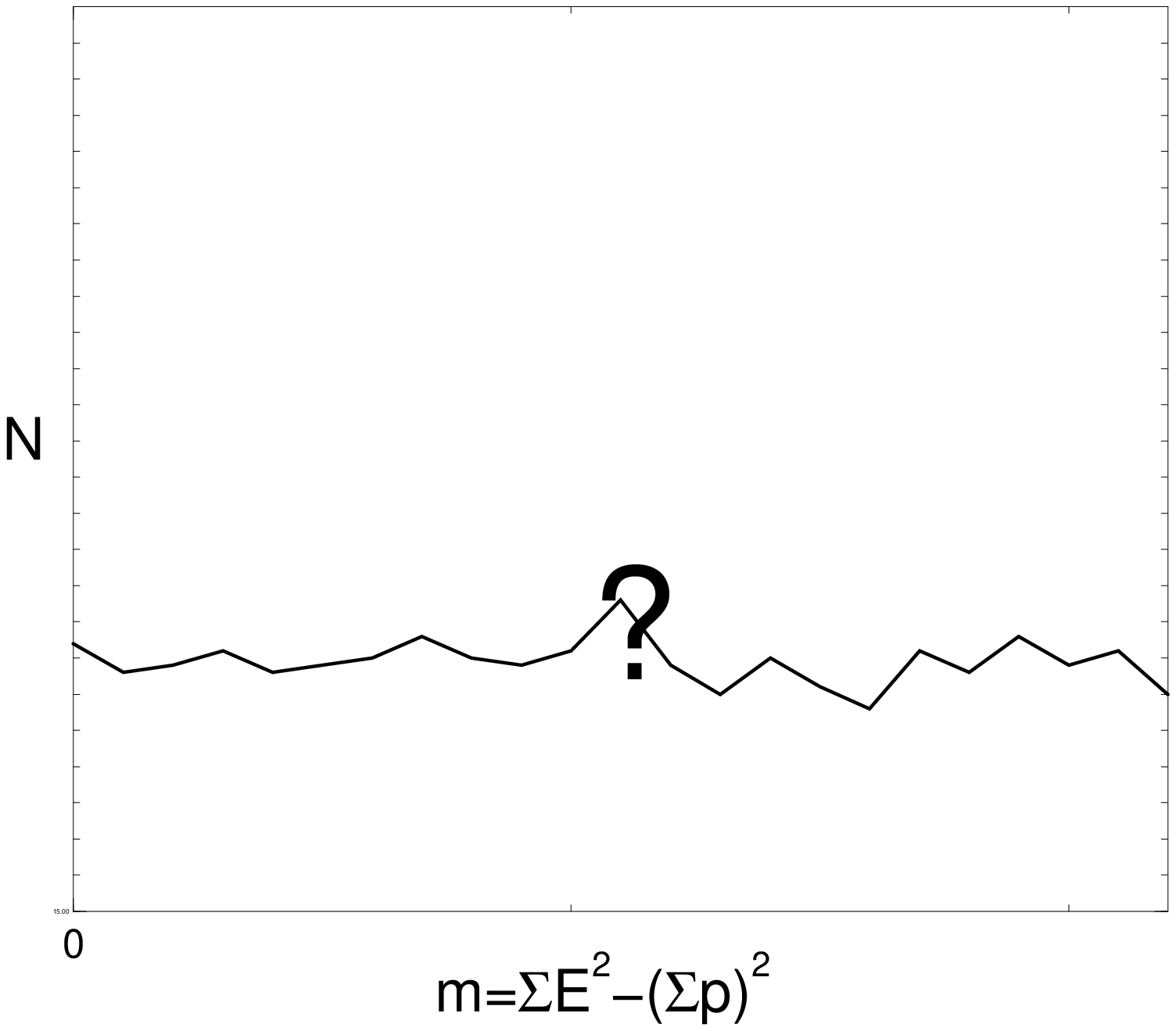}
\caption{Re-scattering can inhibit resonance reconstruction. Left: in case 
the resonance decay products reach the detector without further interactions, 
their invariant mass distribution should yield a clear peak at the
resonance mass, with the yield corresponding to resonance abundance. 
Right: should these decay products undergo re-scattering
before reaching the detector, the signal may be indistinguishable from
the background caused by unrelated particle pairs. Generally a weakening 
by the medium  of the resonance invariant mass signal must be allowed for.
 \label{rsc_fig}}
\end{figure}

A quantitative understanding of the above picture is crucial
for any meaningful analysis of the final state particles.
Many probes of deconfinement are most sensitive when 
the dense hadron matter fireball breakup is sudden and
re-interaction time short or non-existent.
Final state particles could, however also emerge remembering relatively
little about their primordial source, having been subject
to re-scattering in purely hadronic gas phase.
In fact, theoretical arguments have been advanced in support of
both a sudden reaction picture  \cite{Raf00,Let00} and
a long re-interaction timescale \cite{pbm99}.
Both pictures have been applied to phenomenological fits of hyperon
abundances and distributions \cite{PBM01,searchqgp}.\\
\vskip -0.5cm
It is apparent that hyperon resonances can be crucial in resolving this 
ambiguity:       \\
their initial abundance, compared to stable particles
with the same quark composition, depends primarily to the temperature
at hadron formation.
However, the observed  
abundance will be potentially quite different,
and will strongly depend on the lifetime of the interacting hadron gas
phase.
Resonances can only be observed via invariant mass reconstruction, and
their short lifetime means that they can decay within the interacting hadron
gas (Fig. \ref{rsc_fig}).
In this case, the decay products
can undergo re-scattering, and emerge from the fireball with no memory about 
the parent resonance.
Thus, the observable 
resonance abundance is sensitive to precisely those parameters
needed to distinguish between the sudden and gradual freeze-out models.

This paper offers a  pedagogical description of what we know
about resonance production in heavy ion collisions.
We start with a review of presently available experimental data,
and the open questions which arise.
We then proceed to describe how to calculate the initial resonance
abundance and the effect of re-scattering in terms of the hadronization
temperature and the lifetime of the interacting hadron gas.    
We show how these two parameters can be extracted
from the experimental observations. Finally, we discuss
possible answers to experimental challenges raised in the first section, and
suggest ways by which these questions could be resolved by further
measurements. This survey incorporates  recently
published experimental measurements \cite{fri01,fat01,mar01,mar01b,xu01}
and theoretical results \cite{torrieri1,torrieri2,torrieri3}.
\newpage
\begin{figure}[h]
\psfig{width=7.cm,clip=1,figure=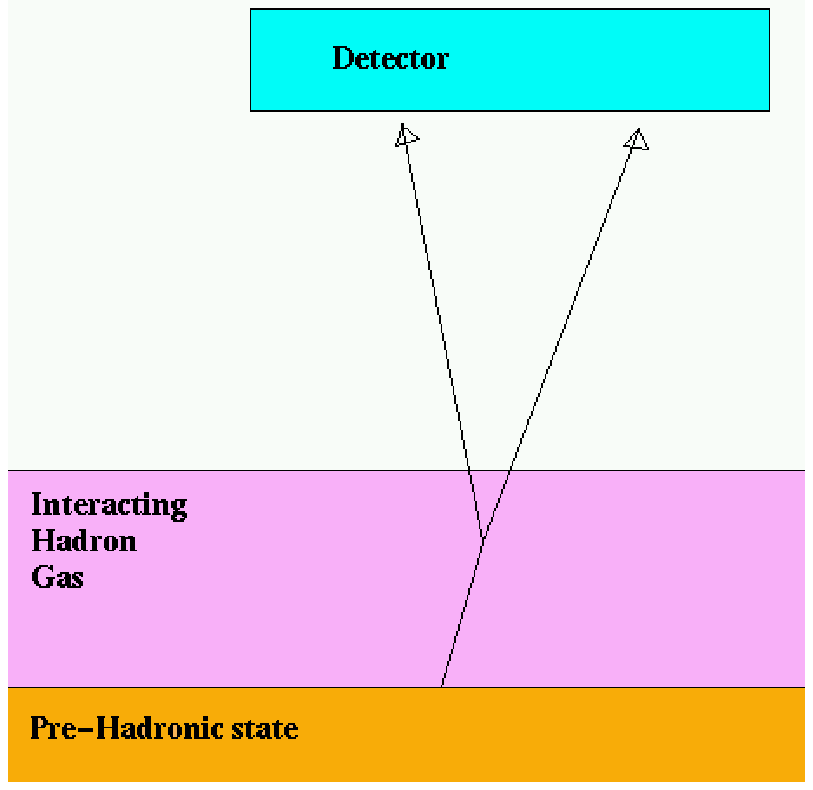}\hspace*{0.4cm}
\psfig{width=7.cm,clip=1,figure=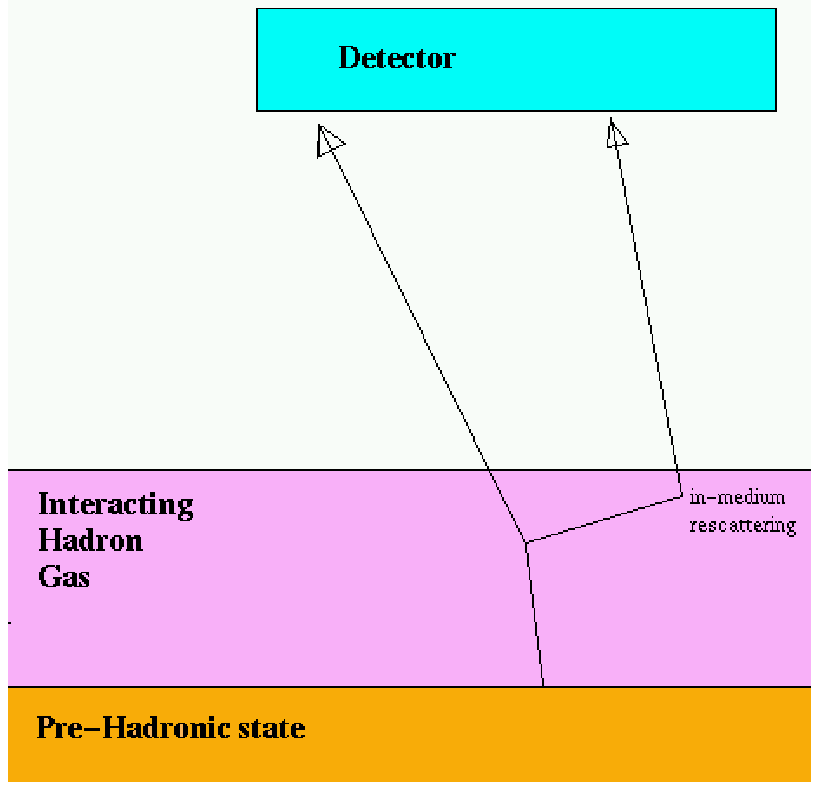}
\end{figure}
\begin{center}
 $\Downarrow$ \color{white}.....................................................\color{black}$\Downarrow$
\end{center}
\section{Experimental measurements}

\subsection{Resonance Reconstruction}

Resonances are detected through invariant mass reconstruction
of their decay products. 
The channels relevant for the here considered $K^{*0},\Lambda(1520) $
 resonances are:

\begin{eqnarray*}
\overline{K}^{*0}(892) \rightarrow \pi^{+} + K^{-},\\
K^{*0}(892) \rightarrow \pi^{-} + K^{+}, \\
\Lambda(1520) \rightarrow p + K^{-}, \\
 \label{kaondecay}
\end{eqnarray*}
and the proposed $\Sigma^{*}$ measurement would be performed  via the
observation of the decay:
\begin{eqnarray*}
\Sigma^*(1385) \rightarrow \Lambda + \pi. 
\end{eqnarray*}
At present, NA49 reported a $\overline{K^*} \rightarrow K^- \pi^+$ signal
\cite{fri01},
while STAR published an average of $K^*$ and $\overline{K^*}$ yields \cite{fat01}.

\begin{figure}[tb]
 \centering
 \includegraphics[width=0.55\textwidth]{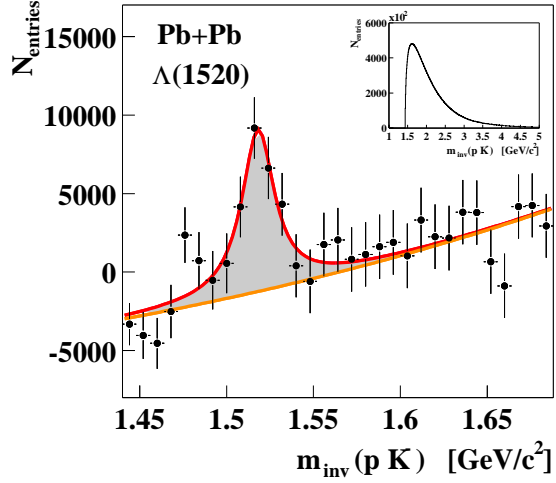}
 \caption{$\Lambda$(1520) mass plot after mixed event
background subtraction with a Breit-Wigner fit from NA49 at
$\sqrt{s} = $ 17 GeV, insert plot: Mass plot before mixed event
background subtraction \cite{mar01}.}
 \label{pbpb}
\end{figure}

Both the STAR \cite{ack99} and the NA49 \cite{afa99} experiments
perform this reconstruction based on data collected in their large volume 
{\em Time Projection Chambers} (TPC) detectors.
Charged particle momenta are obtained by a measurement of their trajectories in a
uniform magnetic field, while particle identification is done evaluating
the rate of energy loss $dE/dx$ in the TPC \cite{Instr}.
A centrality trigger selects the 14\% (STAR) and 5\%
(NA49) most central inelastic interactions. The decay daughter
candidates are selected via their momenta and $dE/dx$. The resonance
signal is obtained by the invariant mass reconstruction of each
pair combination and the subtraction of a mixed event background
estimated by combining candidates from different events (
Figure~\ref{pbpb}). The multiplicity is obtained after applying
correction factors for acceptance, particle identification cut,
reconstruction efficiency and branching ratio.

\subsection{Determination of Width and Mass}

The measured width and mass of $\Lambda$(1520) and K* in heavy ion
collisions is in agreement within the statistical errors with the
values from the Particle Data Group (table~\ref{resowidht}).

\begin{table}
\begin{tabular}{cclllllc}
\hline
  \tablehead{1}{c}{b}{collision \\}
 & \tablehead{1}{c}{b}{Energy \\ GeV}
  & \tablehead{1}{l}{b}{particle \\}
  & \tablehead{1}{l}{b}{mass \\ MeV/c$^{2}$}
& \tablehead{1}{l}{b}{$\sigma _{\mathrm{mass}}$ \\ MeV/c$^{2}$ }
 & \tablehead{1}{l}{b}{width \\ MeV/c$^{2}$}
&  \tablehead{1}{l}{b}{$\sigma _{\mathrm{width}}$ \\ MeV/c$^{2}$}
 & \tablehead{1}{l}{b}{reference \\ }   \\

\hline
Pb+Pb& 17.2 &$\Lambda$(1520) & 1518.1& 2.0 &  22.7  &6.5 & \cite{mar01} \\
Au+Au &130 &K*$^{0}$       &893.0 & 3.0  & 58.0 & 15.0  & \cite{xu01}\\
Au+Au &130 &$\overline{K^{*0}}$  &    896.0 & 4.0 &  63.0  &11.0& \cite{xu01} \\

PDG & &$\Lambda$(1520)  & 1519.5 & 1.0 & 15.6 & 0.6 & \cite{pdg98}\\
PDG & &K*$^{0}$ and $\overline{K^{*0}}$ &  896.1 & 0.27  & 50.7 & 15.0 & \cite{pdg98}\\

\hline
\end{tabular}
\caption{Resonance mass and width}
 \label{resowidht}
\end{table}

Within the statistical error there are no width broadening or
mass shifts observed. While there is no calculation available
that estimates the width and mass profile of a resonance by
taking the medium density evolution during the expansion into
account, shifts in width and mass would be expected in a lengthy hadron
gas phase \cite{aichelin}.
A prediction based on relativistic chiral SU(3) dynamic
calculations \cite{lut01,lut2} gives a 100 MeV broadening of the
$\Lambda$(1520) resonance and a mass shift of about 100 MeV to
lower masses in medium at a density of $\rho$ = 0.17~fm$^{-3}$.

The fact that the $\Lambda$(1520) NA49 signal from Pb+Pb reactions
 shows no width broadening within the errors, indicates that
only the decay products coming from the vacuum $\Lambda$(1520) 
decay are observed. 
Any medium modified $\Lambda$(1520) resonances must thus decay 
rapidly  within the medium, with decay products rescattering, or, 
alternatively no such  $\Lambda$(1520)-medium interaction is present.
To distinguish these alternatives one needs to consider the
yields of $\Lambda$(1520).

\subsection{Particle Multiplicities}

Experimental results for K* and $\Lambda$(1520) resonance
production in heavy ion collisions are available from NA49
experiment in central Pb+Pb and p+p at $\sqrt{s} = $ 17 GeV and
from STAR experiment in central Au+Au $\sqrt{s_{\rm NN}} = $ 130
GeV collision energy and are shown in table~\ref{resonumbers}.\\

\begin{table}[tb]
\begin{tabular}{cclllcc}
\hline
  \tablehead{1}{c}{b}{collision}
 & \tablehead{1}{c}{b}{Energy \\ GeV}
  & \tablehead{1}{l}{b}{particle}
  & \tablehead{1}{l}{b}{yield}
 
 & \tablehead{1}{l}{b}{reference}   \\
 
\hline
 
p+p & 17.2 & $\Lambda$(1520)      & 0.0121 $\pm$ 0.003 &
\cite{fri01,mar01} \\
p+Pb & 17.2 & $\Lambda$(1520)     &  0.0072 $\pm$ 0.003    &
\cite{fri01}\\
Pb+Pb & 17.2 & $\Lambda$(1520)    & 1.45 $\pm$ 0.40    &
\cite{fri01,mar01} \\
Pb+Pb & 17.2 & K* 5-10\% most central & 12.88 $\pm$ 4.53              &
\cite{fri01} \\
Au+Au & 130 & K*  (dN/dy)$_{\mathrm{y}<|0.5|}$                 &  10.0 $\pm$ 0.8 &
\cite{xu01,fat01}\\
Au+Au & 130 &  $\Lambda$(1520)    & $<$ 4.2 \tablenote{at 95\%
confidence level}   & \cite{mar01b} \\
 
\hline
\end{tabular}
\caption{Resonance yields}
 \label{resonumbers}
\end{table}

\begin{figure}
\begin{minipage}[b]{0.5\linewidth}
 \centering
\includegraphics[width=0.95\textwidth]{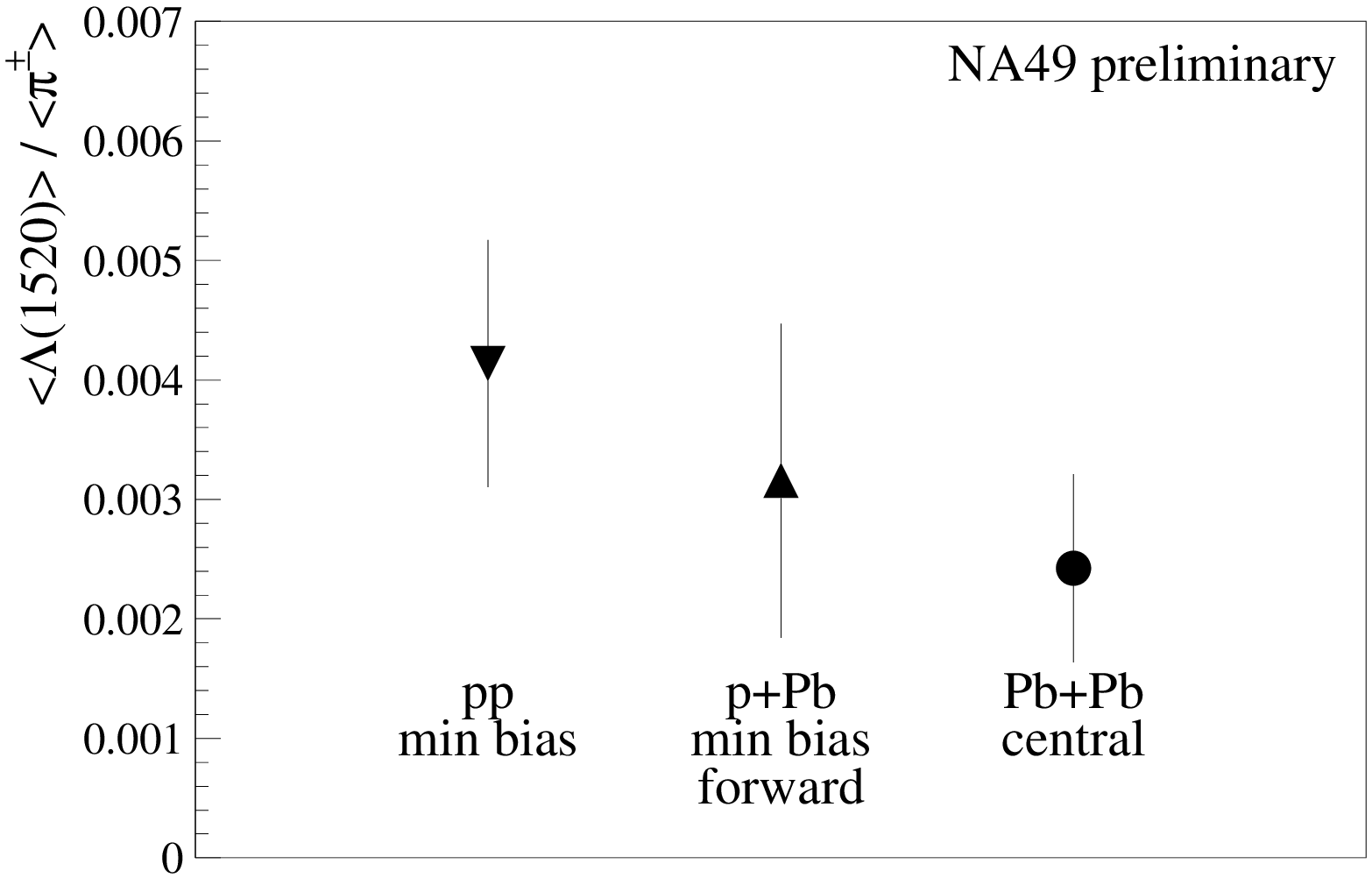}
\vspace{1cm}
 \end{minipage}
 \hspace{-0.5cm}
 \begin{minipage}[b]{0.5\linewidth}
 \centering
 \includegraphics[width=0.85\textwidth]{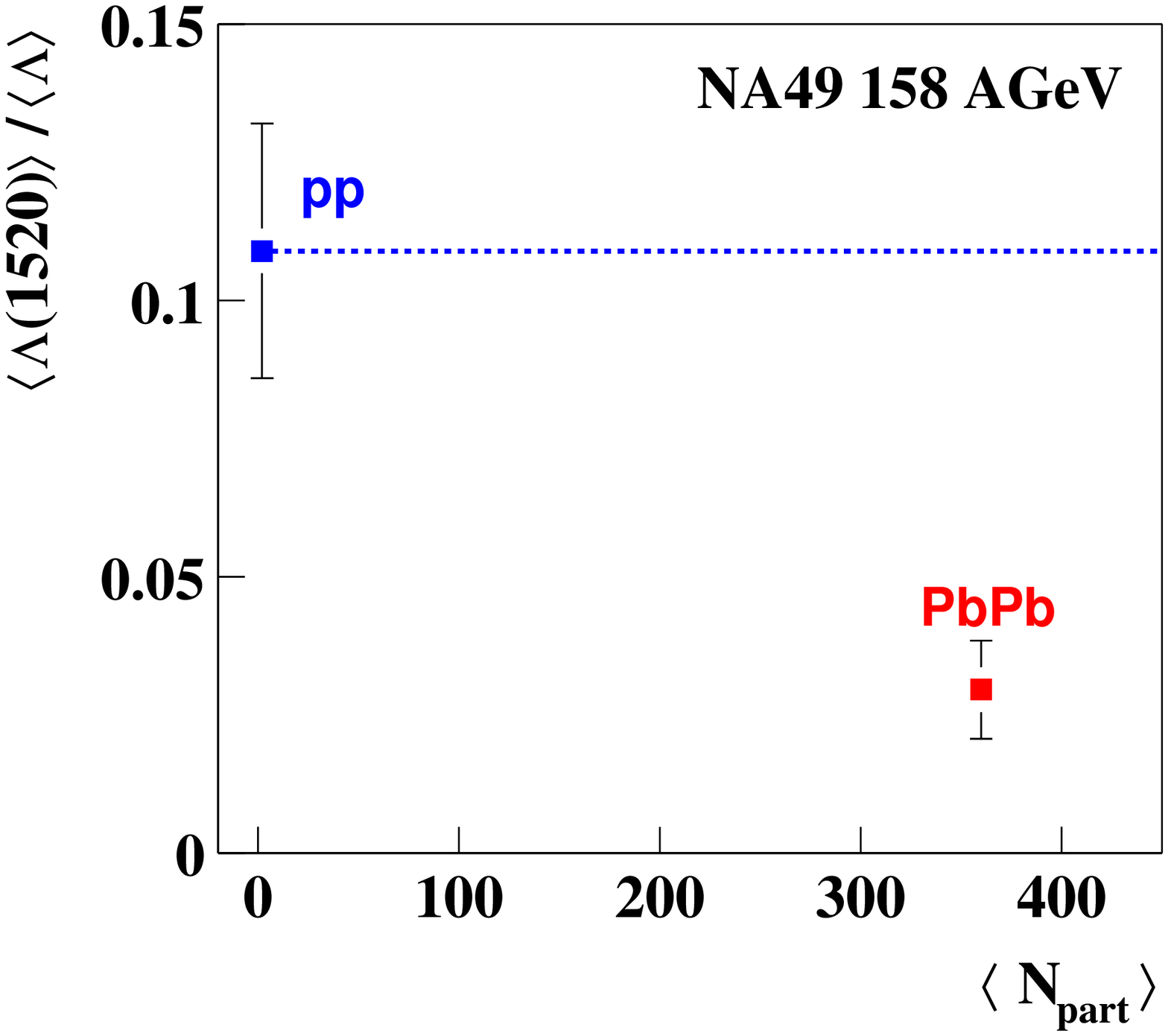}
\end{minipage}
\caption{Left: $\Lambda$(1520)/$\pi^{+}$ as a function of
participants for p+p, p+Pb and Pb+Pb collisions at 158~AGeV
\cite{fri01}. Right: $\langle\Lambda$(1520)$\rangle$/$\Lambda$
ratio as a function of number of participants. Data points are p+p
and central Pb+Pb collisions from NA49.} \label{ratios}
\end{figure}

\begin{figure}[tb]
 \centering
 \includegraphics[width=0.5\textwidth]{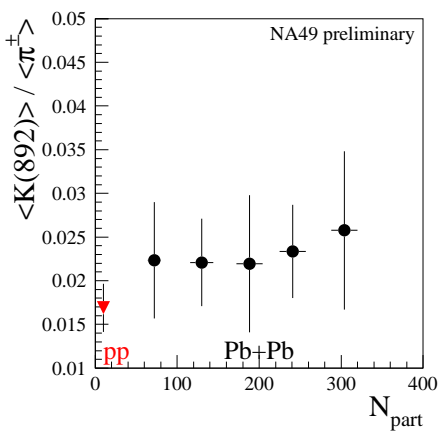}
 \caption{K$^{*}$(892)/$\pi$ as a function of participants for p+p 
and Pb+Pb collisions at 158~AGeV.}
 \label{kstar}
\end{figure}

Comparing $\Lambda(1520) \rightarrow p K$ yields at different
experiments, we see a marked decrease from
elementary p+p to Pb+Pb collisions at
17.2 GeV, which is even more pronounced in the
$\Lambda$(1520)/$\Lambda$ ratio (figure~\ref{ratios} left and
right). This suggests that heavy ion collisions are not merely
superpositions of elementary p+p collisions. Here we can ask the
first question. "Why is the $\Lambda$(1520) signal from the
invariant mass reconstruction in the K-p decay channel
actually suppressed in heavy ion collisions, whereas all
other hyperons are enhanced?''  Moreover, the K* over $\pi$ production at the
same energy indicates no such signal decrease from p+p to Pb+Pb
collisions, shown in figure~\ref{kstar}. The second question is
now "Why is no comparable suppression observed in the $K^*$?''
A hint for the signal loss could be
due to secondary interactions is the measurement of the
$\Lambda$(1520) production in p+Pb collisions which seems to be
decreasing. However the errors are to large to say this conclusively
(figure~\ref{ratios} left).

\begin{figure}[tb]
 \centering
 \includegraphics[width=0.85\textwidth]{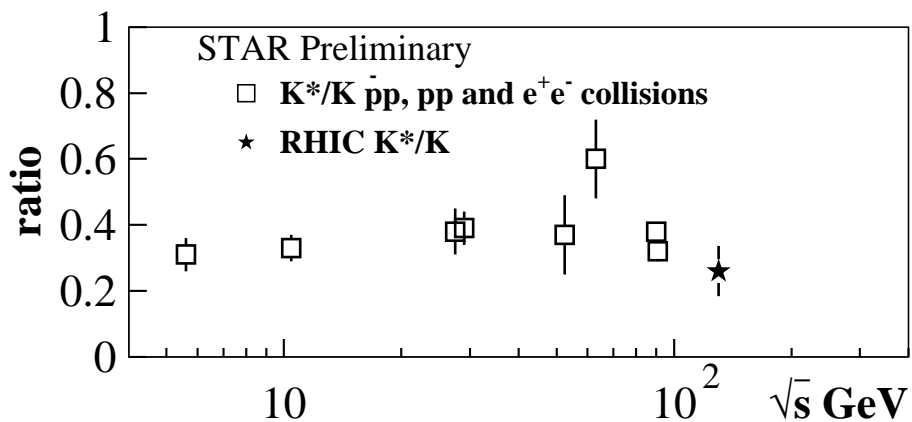}
 \caption{K*$^{0}$/K measured in different colliding systems at
 different energies. For elementary e$^{+}$e$^{-}$ and pp collisions in
 comparison \cite{can79,alb94,agu91,der85,dri81,ake82,abe99,pei96,chl99}
 with the heavy ion collision Au+Au at RHIC energies $\sqrt{s_{\rm NN}} = $ 130
GeV \cite{fat01}. }
 \label{kstarstar}
\end{figure}

For the collision energy of $\sqrt{s_{\rm NN}} = $ 130 GeV there
is only an upper limit for the $\Lambda$(1520), namely
4.2 at 95\% confidence level (2$\sigma$). The expected
multiplicity from extrapolated elementary p+p reactions including
an addition factor of 2 for strangeness enhancement (taken from
$\Lambda$ at SPS \cite{gaz99}) is $\sim$ 7.7. The upper limit
estimate indicates that we see at RHIC energies the same trend of
signal loss as at SPS energies. The K*/K ratio at RHIC energies
fits in within errors to the ratios from elementary p+p and
e$^{+}$+e$^{-}$ collisions figure~\ref{kstarstar}, that leads us
to the conclusion that there is no indication of K* signal loss
related to K yield. Our conclusions from the results from SPS and
RHIC energies are in agreement.

\begin{figure}[tb]
 \centering
 \includegraphics[width=0.65\textwidth]{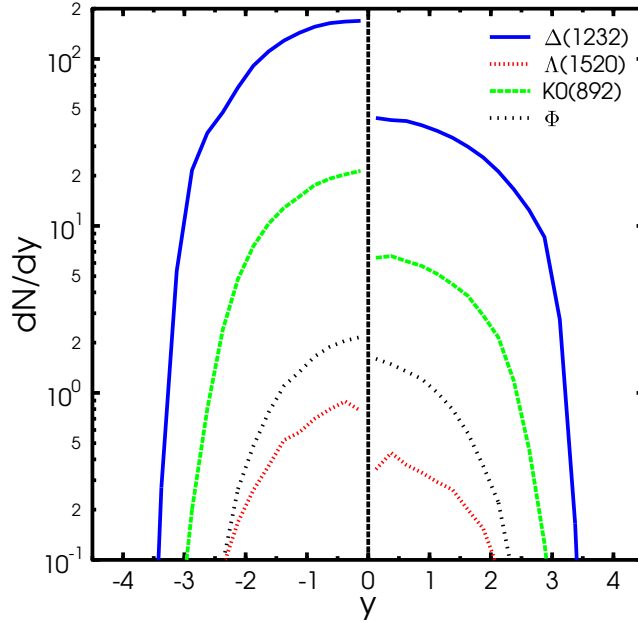}
 \caption{Urqmd calculation. Rapidity distributions of produced
  resonances (left half), and of the resonances which remain observable
after  thermal freezeout\cite{ble02}.}
 \label{urqmd}
\end{figure}

A recent UrQMD calculation \cite{ble02} predicts for the K* a
signal loss of 66~\% and for the $\Lambda$(1520) 50~\% at SPS
energies due to re-scattering of the decay products
(figure~\ref{urqmd}). At RHIC energies the signal loss is on the
order of 55\% for K* and 30~\% for $\Lambda$(1520) \cite{ble02b}.
The K* yield at A+A collisions for SPS and RHIC energies shows 
little or no indication of signal loss which was expected.
 This means
that the mechanism of re-scattering alone can not explain the
signal loss of the $\Lambda$(1520) with respect to the K*
production.

As table \ref{thermalmodels} shows, the agreement of the data with 
thermal model predictions presents a considerable problem as well.
At SPS energies, thermal model calculations over-predict  the
$\Lambda(1520)$ yield by a factor of two ~\cite{bec01},
while the upper limit measurement of
$\Lambda$(1520) from STAR is 30\% lower than the predicted value.
However, published thermal model-calculations significantly under-predict
the $K^*$ (although, as will be seen in the next section, the observed
$K^*$ is consistant with thermal production at a lower freezeout temperature).

\begin{table}[tb]
\begin{tabular}{ccllc}
\hline
  \tablehead{1}{c}{b}{collision}
 & \tablehead{1}{c}{b}{Energy \\ GeV}
  & \tablehead{1}{l}{b}{particle}
  & \tablehead{1}{l}{b}{predicted \\yield}
 & \tablehead{1}{l}{b}{reference}   \\

\hline

Pb+Pb & 17.2 & $\Lambda$(1520)    & 3.48   & \cite{bec98} \\
Pb+Pb & 17.2 & $\Lambda$(1520)    & 5.20   & \cite{pbm99} \\
Au+Au & 130 & $K^*/h^-$  & 0.037 & \cite{bmrhic} \\

\hline
\end{tabular}
\caption{Particle yields from thermal model predictions}
 \label{thermalmodels}
\end{table}

\section{Modeling resonances in heavy ion collisions}

\subsection{Direct production at hadronization}

We assume that at hadronization, a volume element
will be at thermal and chemical equilibrium in its local rest frame.
This means hadrons produced 
directly from a medium at temperature T (much lower
than the particle mass, as in all cases considered here) 
fill the available phase space according to the relativistic Boltzmann 
distribution:
\begin{eqnarray}
\label{direct}
d N \propto g d^3\! p f(\lambda,E,T),\\
f(\lambda,E,T)=  \lambda e^{-E/T}.
\end{eqnarray}
Here $g$ is the statistical degeneracy and $\lambda=e^{\mu/T}$ is
the hadron's fugacity.
Since a hadron is a composite state of flavored
quarks, it is natural to parametrize
its fugacity as a product of the flavor fugacities of each constituent.
As a further step, each quark's fugacity can be decomposed into a
term $\lambda_{\mathrm{i}}$ controlling the net flavor ($q - \overline{q}$) and
an occupancy number $\gamma_{\mathrm{i}}$ which determines the number of 
quark-antiquark pairs. Therefore, 
\begin{eqnarray}
\label{fugacities}
\lambda&=&\prod^{n}_{\mathrm{q=1}} \lambda_{\mathrm{i}} \gamma_{\mathrm{i}},\\
\lambda_{q}&=&\lambda_{\overline{q}}^{-1},\\
\gamma_{q}&=&\gamma_{\overline{q}},
\end{eqnarray}
and the precise values of $\lambda_{\mathrm{i}},\gamma_{\mathrm{i}}$ can be determined 
from global particle abundance fits~\cite{Let00}.

We can now change to rapidity ($y$) and transverse mass ($m_T$) 
coordinate system, where transverse and longitudinal directions are 
defined with respect to the beam.  (The advantage of this system is that $m_T$ is
invariant between the lab and the center of mass frames, while
the rapidity transforms additively).
\begin{eqnarray}
\label{ymt}
dN &=&g m_T^2 \cosh(y) d m_T dy f(\lambda,m_T \cosh(y),T)\\
m_T&=&\sqrt{m^2+p_T^2}\\
y  &=&\frac{1}{2} \ln\left(\frac{E+p_L}{E-p_L}\right)
\end{eqnarray}
To get particle ratios in a certain region of phase space
collective expansion (flow) as well as freeze-out geometry has to be included.
The most general way to do this is to introduce a space-time hypersurface
(a 3-D surface in 4-D space-time) $d \Sigma_{\mu}$, which will transform as
a 4-vector.  This vector can either be time-like, corresponding to
a freeze-out surface evolving in time, or space-like, corresponding
to a freeze-out occurring simultaneously across the fireball volume.
The total number of emitted particles will then be
\begin{equation}
\label{current}
N_{\mathrm{i}}= \int d \Sigma_{\mu} (x) j^{\mu} (x).
\end{equation}
We can describe the current 4-vector $j^{\mu}$ 
in terms of the hadron density in the 
volume element at rest, (see Eq. \ref{direct})
\begin{equation}
\label{statcurr}
j^{\mu}=n_{\mathrm{rest}} u^{\mu} = f(\lambda,E_{\mathrm{local}},T) \frac{p^{\mu}}{E},  
\end{equation} 
where $E_{\mathrm{local}}$ is the energy in the local rest frame with respect
to flow, and  $u^{\mu}$ is the volume element's 4-velocity.

In the limit of a homogeneous fireball (Temperature and chemical
potential constant in space) and full phase space, 
the total number of produced particles is, as expected,
independent of flow and hadronization geometry, since in this case
Eq. (\ref{current}) becomes
\begin{equation}
\label{homo}
N=\int d \Sigma_{\mu} (x) j^{\mu} (x) 
= \int d \Sigma_{\mu} (x) u^{\mu} (x) 
\int d^3 p f(\lambda,E_{\mathrm{local}},T)
= V \int d^3 p f(\lambda,E_{\mathrm{local}},T).
\end{equation}
We can now find N analytically
\begin{equation}
\label{nanal}
N= V \lambda \int  d^3 p e^{-(p^2+m^2)/T}= \frac{1}{2 \
\pi^2} \lambda m^2 T K_{2} (m/T)
\end{equation}
remembering the properties of Bessel functions
\begin{equation}
\label{bessel}
K_n (x)=\frac{2^n n!}{(2n)!}x^{-n} 
\int_{0}^{\infty} d z (z^2+x^2)^{-\frac{1}{2}} z^{2 n} e^{-\sqrt{z^2+x^2}},
\end{equation}
If the fireball is not homogeneous, or we want to only calculate the yield
within the experimental acceptance limits, we need to use a more
general result.
Putting Eqs. (\ref{current}) and (\ref{statcurr}) together one gets
the relativistically invariant Cooper-Frye formula \cite{cooperfrye1}
\begin{equation}
\label{cooperfrye}
E \frac{d^3 N}{d^3 p} = \frac{1}{2 \pi m_T}\frac{d N}{d y m_T}
= \int d \Sigma_{\mu} p^{\mu} f(\lambda,p_{\mu} u^{\mu},T) 
\theta(\Sigma_{\mu} p^{\mu}),
\end{equation}
where the step function $\theta(\Sigma_{\mu} p^{\mu})$ has been introduced
to eliminate unphysical emission in the direction opposite to expansion
\cite{cooperfrye3,cooperfrye2}.
Eq. (\ref{cooperfrye}) can be integrated to yield the total particle number over
an arbitrary phase space region
\begin{equation}
\label{totnumber}
N= \int \frac{d^3 p}{E} \int  d \Sigma_{\mu} p^{\mu}
 f(\lambda,p_{\mu} u^{\mu},T) \theta(\Sigma_{\mu} p^{\mu})
\end{equation}
Many of the parameters used in these formulae can be eliminated by judicious
choice of observables.
For instance, the ratio of a hyperon resonance to its
ground state is independent of the chemical potential, since
all chemical potential terms cancel out.
Suitable ratios include $\Sigma^{*}/\Lambda$,  $\Lambda(1520)/\Lambda$,
$K^{*0}/K^{+}$, $\overline{K^{*0}}/K^{-}$.  
A particular case is the $K^{0}/\overline{K^0}$ system:
While the $K^{*0}$ decays long before it has a chance of
oscillate, the experimentally observed $K_{S}$ is a superposition of
$K^0$ and $\overline{K^0}$.   Suitable average are 
$\overline{K^{*0}}/K^{-}$ (measured by NA49 \cite{fri01} ) $K^{*0}/K^{+}$
and $(K^{*}+\overline{K^{*}})/(K^{+}+K^{-})$ (measured by STAR \cite{fat01}).

Moreover, we found that for these ratios (more generally for
ratios of particles with comparable masses), 
effects due to flow and surface geometry
cancel out to a very good approximation and Eq. (\ref{nanal}) gives the produced
particle ratios to a good accuracy.

\subsection{Feed down from resonance decays}

Many of the hyperons considered here can be produced either directly at
hadronization or in the decay of a resonance produced earlier.
For instance, 
\begin{equation}
\label{1520ratio}
\frac{\Lambda(1520)}{\Lambda_{\mathrm{total}}}
=\frac{\Lambda(1520)}
{\Lambda+(\Sigma^{0}\rightarrow \Lambda \gamma)+(\Sigma^* \rightarrow \Lambda \pi) 
+ (\Sigma^{*} \rightarrow \Sigma^0 \pi \rightarrow \Lambda \pi \gamma)}.
\end{equation}
Unless a resonance will be directly reconstructed by the experiment, its
decay products will be a component in the hyperon abundances (albeit
with a subtly different momentum distribution).   This contribution, therefore,
will need to be taken into account through calculation.
As in the case of direct production, all kinematic effects are
integrated out in a calculation of the total yield,
and one simply sums each term in the resonance contribution
as a term of the form given in Eq. (\ref{nanal}).
If a finite acceptance in phase space needs to be taken into account,
however, a distribution of decay products will need to be computed from
the statistical distribution of resonances.

We assume that, in a decay of the form,
\begin{equation}
R \rightarrow 1 + 2+...,
\label{dectype}
\end{equation}
any dynamical effects in the decay (the S-matrix) average out
over a statistical sample of many resonances.
In other words, in the rest frame comoving with the 
``average'' resonance, the distribution
of the decay products will be isotropic.
The rate of particles of type 1 (as in Eq. (\ref{dectype}))
produced with momentum $\vec{p^{*}_{1}}$ in the frame
at rest w.r.t. the resonance will
then simply be given by  the Lorenz-Invariant phase space factor
of a particle of mass $M_1$ and momentum $\vec{p^{*}_{1}}$ within a system
with center of mass energy equal to the resonance mass $M_R$.
\begin{equation}
\label{decayphase}
\frac{d^3 N_1}{d^3 p^{*}_{1}}=b \int \prod_{i=2}^{n}
 \frac{d^3 p^*_i}{2 E^*_i} \delta(\sum_{i=2}^N p^*_i-p^*_1) 
\delta(\sum_{i=2}^N E^*_i-E^*_{1}-M_R),
\end{equation}
where $b$ is the branching ratio of the considered decay channel.
All that is left is to change coordinates from the resonance's 
rest frame $(p^*,\,E^*)$ to
the lab frame $(p,E)$.

If more than two bodies are considered, this calculation becomes
more involved \cite{phasespace}.
For the general N-body case, it is better left to Monte Carlo methods
\cite{mambo}.
In the case of the 2-body decay, the situation is greatly
simplified by the fact that $p^*,\,E^*$ are fixed by energy-momentum 
conservation  to,
\begin{eqnarray}
\label{2bodpstar}
E^{*}_1=\frac{1}{2M_R}(M_R^2-m_1^2-m_2^2),\\ 
p^{*}_1=-p^{*}_2=\sqrt{E_1^{*2}-m_1^2}.
\end{eqnarray}
Putting the constraints in Eq. (\ref{2bodpstar}) into Eq. (\ref{decayphase})
one gets, after some algebra
\cite{resonances}
\begin{eqnarray}
\label{reso}
\frac{dN}{d {m^2_{T1}} d y_1 } &=&
\frac{ b}{4 \pi p^{*}_1}
\int_{Y-}^{Y+} dY_1
\int_{M_{T}-}^{M_{T}+} dM_{T1}^{2} J 
\frac{d^2 N_{R}}{dM_{TR}^{2} dY_R} , \\
\noindent J&=&\frac{M_R}{\sqrt{P_{TR}^2 p_{T1}^2 
-(M_R E^{*}_R - M_{TR} m_{T1} \cosh \Delta Y)^2 }},\\
\Delta Y&=&Y_R-y_1.
\end{eqnarray}
$J$ is simply the Jacobian of the transformation from the resonance rest
frame to the lab frame, and the limits of the kinematically allowed integration
region are:

\[\ Y_{\pm}=y_1 \pm \sinh^{-1} \left( \frac{p^{*}_1}{m_{T1}} \right), \]
\[\ M_{T}^{\pm}=M_R 
\frac{E^{*}_R m_{T1} \cosh(\Delta Y) \pm p_{T1} 
\sqrt{p^{*2}_1-m_{T1}^{2} \sinh^{2} (\Delta Y)}}
{m_{T1}^{2} \sinh^{2} (\Delta Y)+m^{2}_1}.\]

We now apply the methods outlined here to calculate ratios at hadronization
for a range of experimentally observable resonances, both at full
and central rapidity ($|y| \le 0.5$).
The total ratio was calculated using the analytical formula in
Eq. (\ref{nanal}), while for mid-rapidity equations in Eq. (\ref{totnumber}) 
and (\ref{reso}) were used.
In the midrapidity calculation we used a 
uniform spherically symmetric expanding fireball, with a hadronization
surface used earlier by fits in \cite{searchqgp} and \cite{resonances},
\begin{equation}
  d \Sigma_{\mu} = (1,- \frac{\partial t_f}{\partial r} \vec{e_r}) d^3 \vec{r},
\end{equation}
where $\vec{e_r}$ is the unit vector in the radial direction and 
$\partial t_f/ \partial r$ is a constant.
We mentioned earlier that flow and
hadronization surface details cancel out to a very good approximation when
hadron masses are comparable.   In fact, when we varied 
$\partial t_f/ \partial r$, the results did not change
by more than a percentage point.
Table \ref{processes} summarizes the decay processes 
considered in our analysis and their parameters (Clebsh-Gordon 
coefficients have been used to estimate decays such as
($N^{*0} \rightarrow N^+ \pi^-)/(N^{*0} \rightarrow N^0 \pi^0)$).

\begin{table}[tb]
\caption{Resonances contributing to $\Lambda$ and K production, with their
degeneracies, rest-frame momentum ($p^*$) and possibility for experimental
reconstruction}
\begin{tabular}{ccccc} 
\hline
\tablehead{1}{c}{b}{g} & \tablehead{1}{c}{b}{Reaction} &
 \tablehead{1}{c}{b}{$p^{*}$\\ MeV}  & 
\tablehead{1}{c}{b}{branching} & \tablehead{1}{c}{b}{visible?} \\ 
\hline
$\approx 4$ & $\Sigma^{* 0}(1385) \rightarrow \Sigma^{0} \pi^{ 0}$ 
& 127 & $ \approx 4 \%$   & No
\\  \hline  
8 & $\Sigma^{* \pm}(1385) \rightarrow \Lambda \pi^{\pm}$ &
208 & $88 \%$  & Yes  
\\  \hline  
4 & $\Sigma^{* 0}(1385) \rightarrow \Lambda \pi^{0}$ &
208 & $88 \%$   & No
\\  \hline  
2 & $\Sigma^{0} \rightarrow \Lambda \gamma$ & 74  & $100 \%$ & No
\\  \hline  
4 & $\Lambda(1520) \rightarrow N \overline{K}$ & 244 & $45 \%$ & Yes
\\ \hline \hline
3 & $K^{*\,0}(892) \rightarrow K^+ \pi^-$ & 291  & $67 \%$ & Yes \\ \hline
\end{tabular}
\label{processes}
\end{table}

\begin{figure}[tb]
\vspace*{-2cm}
\hspace*{.1cm}
\psfig{width=10cm,clip=1,figure=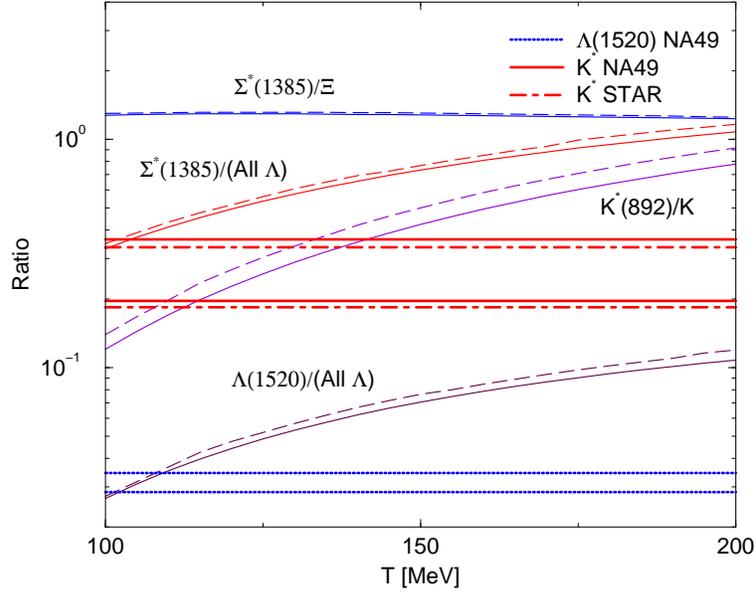   }
\vspace*{-0.21cm}
\caption{ 
Temperature dependence of ratios of $\Sigma^{*}$, $K^{*\,0}$
and $\Lambda(1520)$ to the total number of observed $K$
$\Lambda$ s and $\Xi$ s. Branching ratios are included.
Dashed lines show the result for a measurement
 at central rapidity $\Delta y=\pm0.5$.
The experimental measurements (horizontal lines, not  $\Sigma^{*}/\Xi$) 
included in the diagram were presented in the first part of this paper.
For the $\Xi$ yields in the diagram, chemical potentials were taken from \cite{Let00}.
It should be noted that the $K^*/K$ ratio is actually $\overline{K^*}/K^-$ for
the NA49 measurement, and an average $(K^*+\overline{K^*})/(K^{+}+K^{-})$
for STAR.\label{prodratios}}
\end{figure}

In Fig. \ref{prodratios} we show the relative thermal  production ratios
at chemical freeze-out  over the entire spectrum
of rapidity and $m_{T}$  (solid lines) and central rapidity range 
(dashed lines). 
The sensitivity of resonance yields to hadronization
temperature is apparent for all resonances under consideration.
In particular, the $\Sigma^*$ emerges as a very promising candidate
for further study.
For example,  at the lowest current estimates ($T\simeq 100$ MeV)
of the final break up temperature in 158$A$ GeV SPS collisions 
$33 \%$ of $\Lambda$  are arising from  primary $\Sigma^{*}$, the percentage
rises to slightly more than $50 \%$ if chemical freeze-out
occurs at $T= 190$ MeV.  Both SPS and STAR detectors are capable
to measure hyperon and resonance yields well within the precision
required to distinguish between the two limiting cases.

as we discussed earlier,  the experimentally observed $K^*$ yield is
compatible with the thermally produced ratio for a range of hadronization temperatures,
but the $\Lambda(1520)$ seems very suppressed.
To discuss this further, an estimation of the effect of rescattering
on resonance abundances is necessary.

\vskip -0.5cm

\section{Rescattering}

As explained in the introduction, direct observation of resonances relies
on invariant mass reconstruction.
Therefore, to calculate the observed resonance abundances, 
rescattering after hadronization will have to be taken into account.
This can be looked at within a microscopic model
of hadronic matter such as UrQMD.
As we have shown earlier, UrQMD alone did not explain consistently the experimental
data if  a long re-interaction period and initial thermal model yields were
assumed.  However, as yet there was no detailed
study of the dependence of the resonance yields on the lifetime of the 
interacting hadron gas phase (the interval between chemical and thermal
freeze-out) as well as the chemical freeze-out temperature.   

Here, we present such a study using a ``back of the envelope'' model which 
nevertheless seems to provide an acceptable qualitative description of
the propagation of resonances and their decay products through opaque matter.

We first note in Fig. \ref{prodratios}, that the 
relative $\Sigma^{*}/\Xi$ signal is remarkably independent
(within  5\%) of Temperature.    This is because 
the $\Xi^*(1530)$ contribution cancels nearly exactly the thermal
suppression of the $\Xi$ originating in the $\Xi-\Sigma^{*}$ mass difference.
This effect  could be used for a direct estimate of the
$\Sigma^{*}$ lost through rescattering,
 even without knowledge of the freeze-out 
temperature, should  the chemical
parameters ($\lambda_{\mathrm{i}}$ and $\gamma_{\mathrm{i}}$) be
independently known. A simple test of sudden hadronization model 
consists in measurement of the ratio $\Sigma^{*}/\Xi$.
If it is significantly smaller than unity,
we should expect a re-equilibration mechanism to be present. Otherwise 
sudden hadronization probably applies, since $\Sigma^*$ emerge from
chemical freeze-out without undergoing interactions. 

\begin{figure}[tb]
\includegraphics[width=0.484\textwidth]{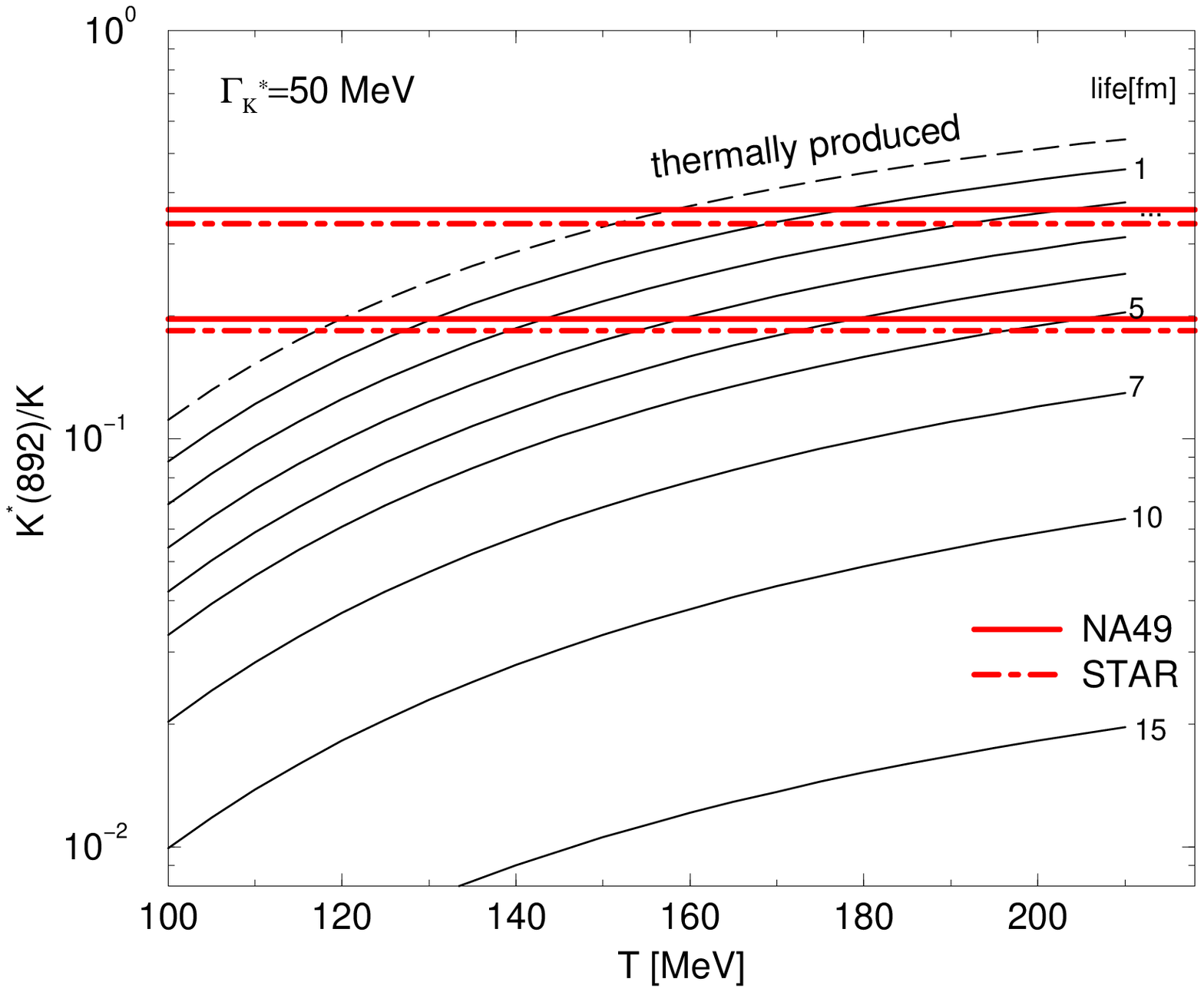}
\includegraphics[width=0.50\textwidth]{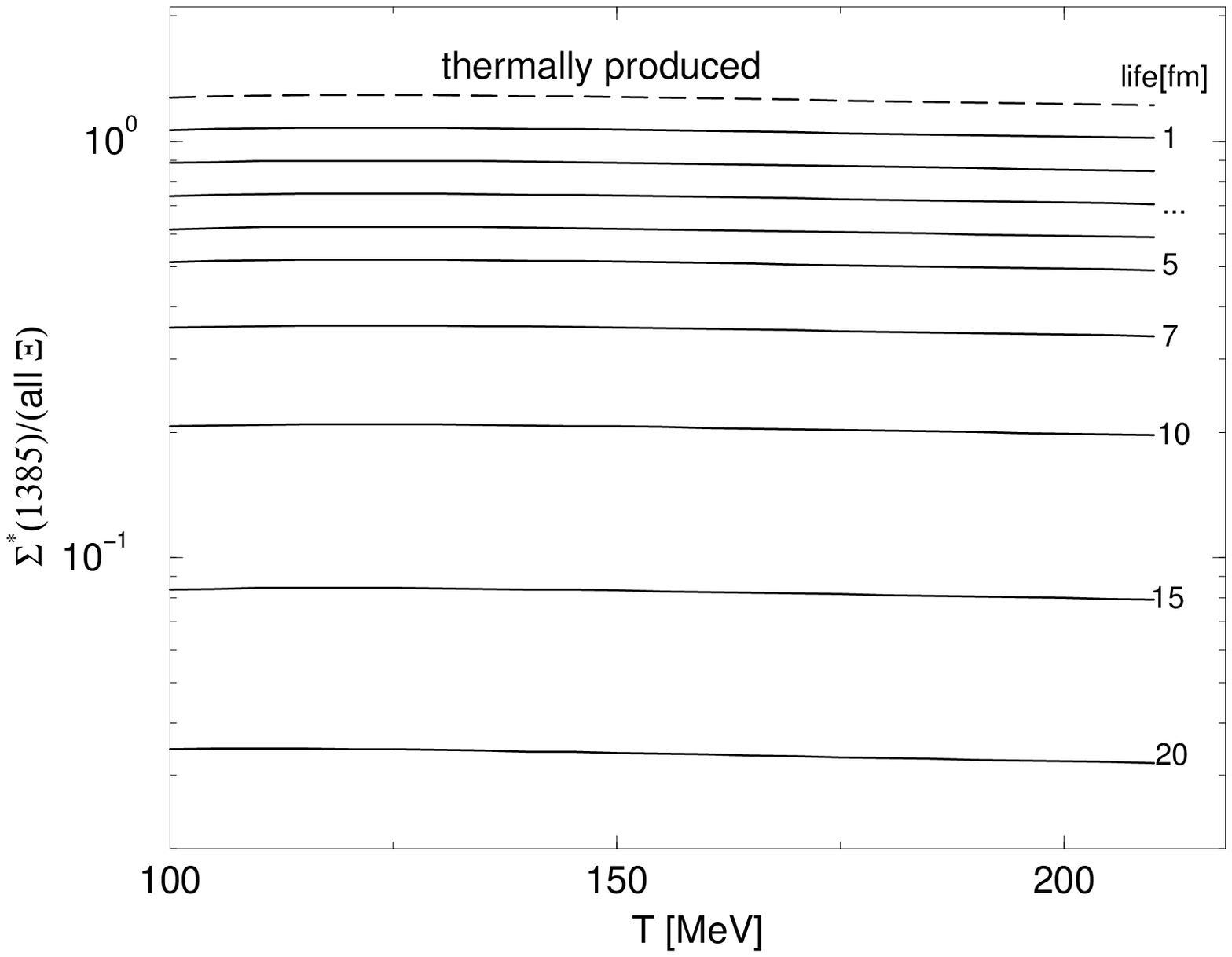}
\caption{Produced (dashed line) and observable (solid lines) ratios
$\Lambda(1520)$/(total $\Lambda$),$\Sigma^{*}$/(total $\Lambda$),
$K^{*0}/K$ and $\Xi$/(total $\Lambda$).
The solid lines correspond to evolution after chemical freeze-out of
1,2,3,4,5,7,10,15,20 fm/c, respectively.  The values at time zero (chemical
freezeout) were taken from
Fig. \ref{prodratios}.
See Fig. \ref{prodratios} caption for the meaning of $K,K^*$.}

\label{obsratios}
\vspace*{-0.9cm}
\end{figure}

We can, however, go further and use  the suppression
of the considered resonances  as a tool capable
of yielding a quantitative estimate of the lifetime of the
interacting hadron gas phase.
We consider the decay of a generic resonance $Y^{*}$
\begin{equation}
\label{decay}
Y^{*} \rightarrow Y \pi \,,
\end{equation}
in a gas of pions and nucleons.
We shall assume that one interaction of either Y or $\pi$ 
is sufficient for that resonance to be undetectable, and that
the decay products travel through the medium with speed $\mathrm{v}_{\mathrm{i}}$
(where $i$ can mean either $Y$ or $\pi$).
The interaction probability is proportional to $\mathrm{v}_{\mathrm{i}}$, the interaction
cross-section of the decay product
with each particle in the hadronic medium 
($\sigma_{\mathrm{i j}}(\mathrm{v}_{\mathrm{i}})$, where $j$ can
refer to either pions, Kaons, nucleons or antinucleons. 
Note that the cross-section itself depends, in a generally 
complicated way, on the incident momentum, and
hence on $\mathrm{v}_{\mathrm{i}}$),
and the particle density in the fireball $\rho_j$.
$\rho_j$ is increased 
by a factor $\gamma_{\mathrm{i}}=1/\sqrt{1-\mathrm{v}_{\mathrm{i}}^2}$
 due to Lorentz-contraction, and decreases
as time passes because of the fireball's collective expansion (parametrized by the
flow velocity $\mathrm{v_{flow}}$, assumed to be of the order of the
relativistic sound speed $c/\sqrt{3}$.)
The time dependence of the densities will therefore be,

\begin{equation}
\label{expansion}
\rho_{j}(t)=\gamma_{\mathrm{i}} \rho_{0j} 
\left( \frac{R}{R+\mathrm{v_{flow}} t}\right)^3,
\end{equation}
and $\rho_{0j}$, the density of j at hadronization, can be calculated
from the chemical freezeout temperatures and chemical potentials.
Putting everything together, the rescattering reaction rate  is
\begin{equation}
\label{scatter}
P_{\mathrm{i}} = \sum_{\mathrm{v_i}}
\left[ \sigma_{i \pi}(\mathrm{v_i}) \rho_{0 \pi}
+\sigma_{i K}(\mathrm{v_i}) \rho_{0 K}
+\sigma_{i N} (\mathrm{v_i}) \rho_{0 N}
 +\sigma_{i \overline{N}}(\mathrm{v_i}) 
 \rho_{0 \overline{N}}) \right] (\gamma \mathrm{v})_{\mathrm{i}}
\left( \frac{R}{R+\mathrm{v_{flow}} t} \right)^3,
\end{equation}
If we use the average,
\begin{eqnarray}
\label{approx}
\sum_{\mathrm{v_i}} \sigma (\mathrm{v_i}) \mathrm{v_{i}} \gamma_{\mathrm{i}}  
\simeq <\sigma><\gamma_{\mathrm{i}} \mathrm{v_i}>
=<\sigma> \frac{p_{\mathrm{i}}}{m_{\mathrm{{i}}}},
\end{eqnarray}
(where p and m are the resonance's momentum and mass)
Eq. (\ref{scatter}) becomes,
\begin{equation}
\label{simplescatter}
P_{\mathrm{i}} = \left[
\left< \sigma_{i \pi} \right>  \rho_{0 \pi}
+\left<\sigma_{i K}\right> \rho_{0 K}+\left< \sigma_{i N} \right> 
\rho_{0N} + \left< \sigma_{i \overline{N}} \right>
\rho_{0 \overline{N}}) \right]  \frac{p_{\mathrm{i}}}{m_{\mathrm{{i}}}}
\left( \frac{R}{R+\mathrm{v_{flow}} t} \right)^3,
\end{equation}
Neglecting in-medium resonance regeneration and particle escape from the 
fireball, the population equation describing the scattering loss abundance
($N_{\mathrm{i}}$)  is:
\begin{eqnarray}
 \frac{d N_{\mathrm{i}}}{d t}&=& \frac{1}{\tau}
 N_{N^{*}} -N_{\mathrm{i}} P_{\mathrm{i}} \,, \quad i=1,2 \\ 
 \frac{d N_{N^*}}{d t} &=& -\frac{1}{\tau} N_{N^{*}}   \,    
 \label{model}
\end{eqnarray}
The required nucleon and antinucleon density at hadronization, 
$\rho_{0N}$ is obtained through Eq. (\ref{nanal}) 
\begin{equation}\label{relboltz}
\rho_{0N}=\frac{g}{(2 \pi \hbar c)^3} 4 \pi m^2 
(\lambda_q \gamma_q)^3 T K_2 (\frac{m}{T}) \,,
\end{equation}
We consider the nucleons to have a mass of $\simeq$1 GeV, and a degeneracy of
six, to take the p,n and the thermally suppressed but higher degeneracy
$\Delta$ contributions into account.
the pion density at hadronization is computed in the massless particle limit, leading to
\begin{equation}\label{pidens}
\rho_{0 \pi}= \frac{\pi^2}{90} T^3\,,
\end{equation}
The model presented here is remarkably insensitive to the individual 
cross-sectional areas.
The values we used in the calculation are given in table
\ref{parameters},but order-of-magnitude variations of the more uncertain
cross-sections did not produce variations of more than $30 \%$.
Similarly, the value of the initial fireball radius $R_0$ (which is constrained
by the entropy per baryon) does not
significantly affect the final ratios.
This reassures us that had we used a more exact approach than
the approximations in Eq. (\ref{approx}), the qualitative features of our
model would not have changed.
The results, however, exhibit a very strong dependence on both the Temperature 
(which fixes the initial resonance yield as well as the hadron
density of the fireball) and fireball lifetime (in a short-lived fireball
not many resonances decay, so their products do not get a chance to rescatter).
Figure \ref{obsratios} shows the dependence of the $\Lambda(1520)/\Lambda$,
$\Sigma^{*}/\Lambda$ and $K^{*\,0}(892)/K$
\begin{table}
\centering
\caption{Scattering model parameters}
\begin{tabular}{cccccc} 
\hline
$\sigma_{\pi N}$ (mb) & $\sigma_{K N}$ & $\sigma_{\pi \pi}$ &
$\sigma_{\pi K}$ &  $\sigma_{N N}$ & $\sigma_{\overline{N} N}$  \\ \hline
24 & 20 & $40$ & $20$ & 24 & 50 \\ \hline\hline
\multicolumn{2}{c} {$\Gamma_{\Sigma^{*}}$ } & 
\multicolumn{2}{c} {$\Gamma_{\Lambda(1520)}$}    &
\multicolumn{2}{c} {$\Gamma_{K^{*0}(892)}$}       \\ \hline
\multicolumn{2}{c} {35 MeV}& 
\multicolumn{2}{c} {15.6 MeV} & 
\multicolumn{2}{c} {50 MeV} \\ \hline\hline
\multicolumn{3}{c} \ \hfill{escape rate (fm$^{-3}$)}\hfill $\big|$ &  
\multicolumn{3}{c} {negligible}   \\ \hline
\multicolumn{3}{c} \ \hfill{v}\hfill $\big|$          & 
\multicolumn{3}{c} {0.5}  \\ \hline
\multicolumn{3}{c} \ \hfill{R(fm)}\hfill $\big|$      & 
\multicolumn{3}{c} {8$\cdot$145/$T$ [MeV]}   \\ \hline
\multicolumn{3}{c} \ \hfill{$\mu_b$}\hfill $\big|$ & 
\multicolumn{3}{c} {$220$ MeV}   \\ \hline
\end{tabular}
\label{parameters}
\end{table}
on the temperature and lifetime of the  interacting phase.
It is clear that, given a determination of the respective signals
to a reasonable precision, a qualitative distinction 
between the high temperature chemical freeze-out scenario 
followed by a rescattering phase and the low
temperature sudden hadronization scenario can be made. We also note that
despite the shorter lifetime of the $\Sigma^{*}$ and
higher pion interaction cross section, 
more $\Sigma^{*}$ decay products should be reconstructible
than in the $\Lambda(1520)$ case, at all but the highest temperatures
under consideration.   
\begin{figure}[tb]
\includegraphics[width=0.49\textwidth]{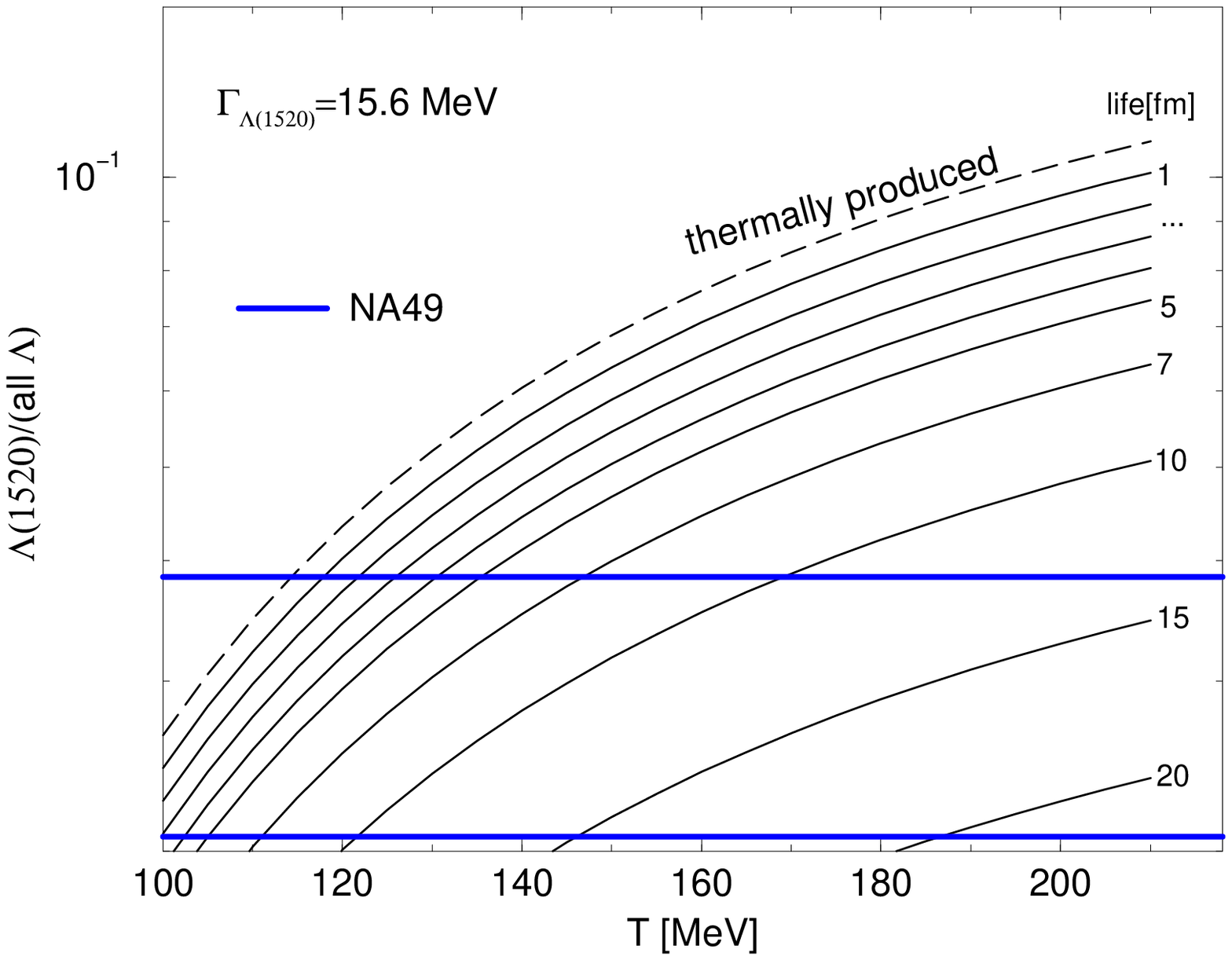}
\includegraphics[width=0.49\textwidth]{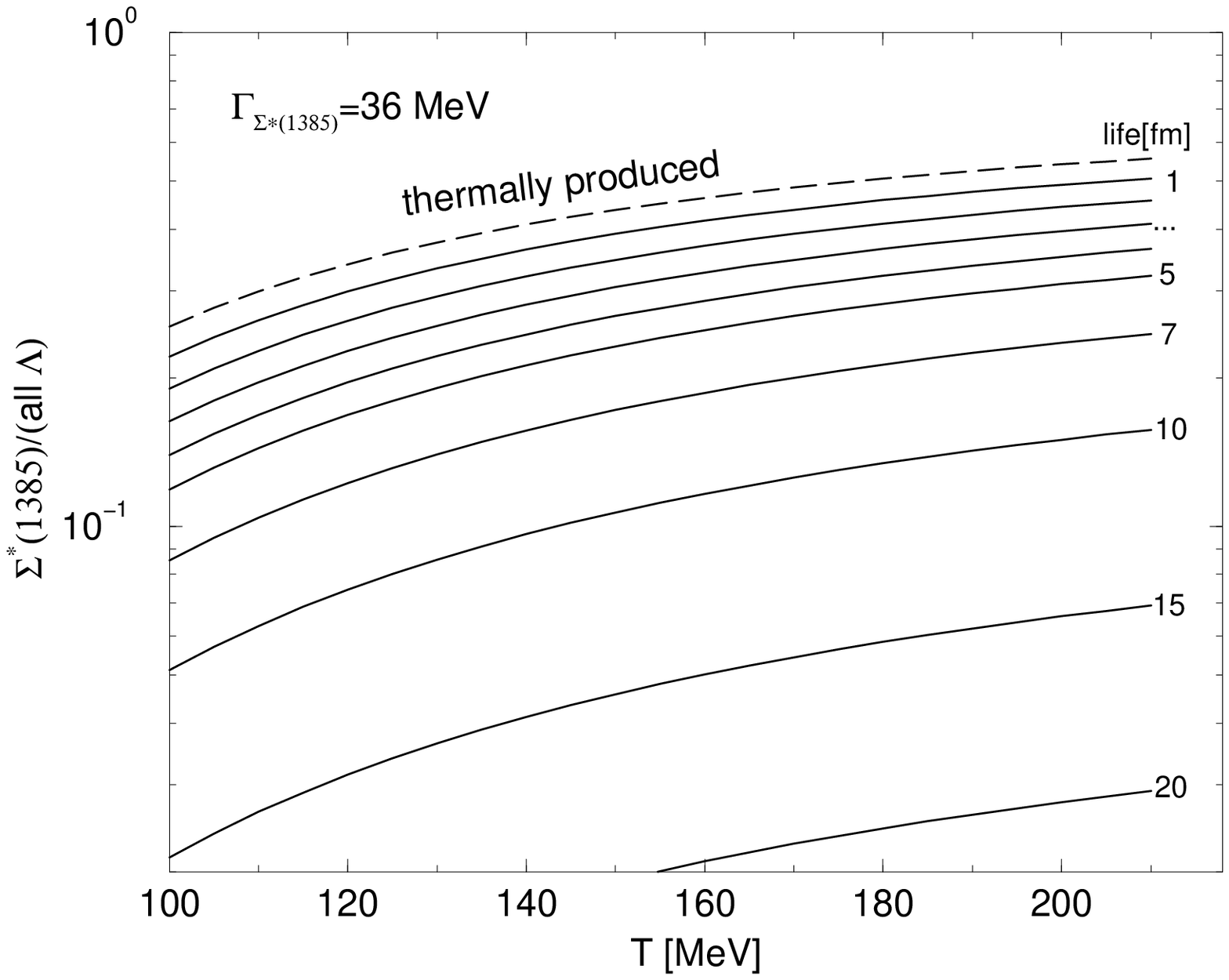}
\end{figure}
This reinforces our proposal that the
$\Sigma^*$ is a very good candidate for further measurement.

\begin{figure}
\psfig{width=7.4cm,clip=1,figure= 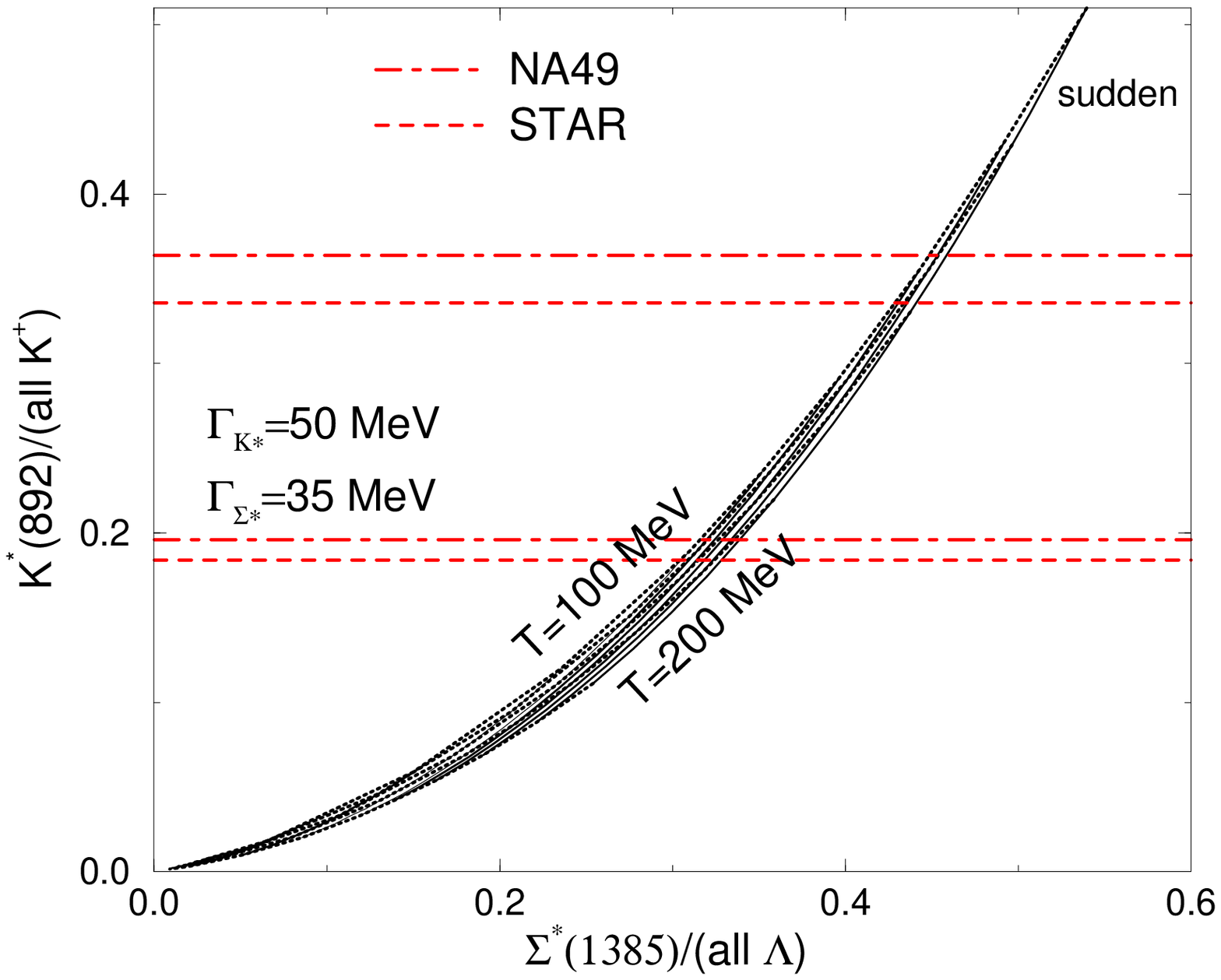 }
\psfig{width=7.2cm,clip=1,figure= 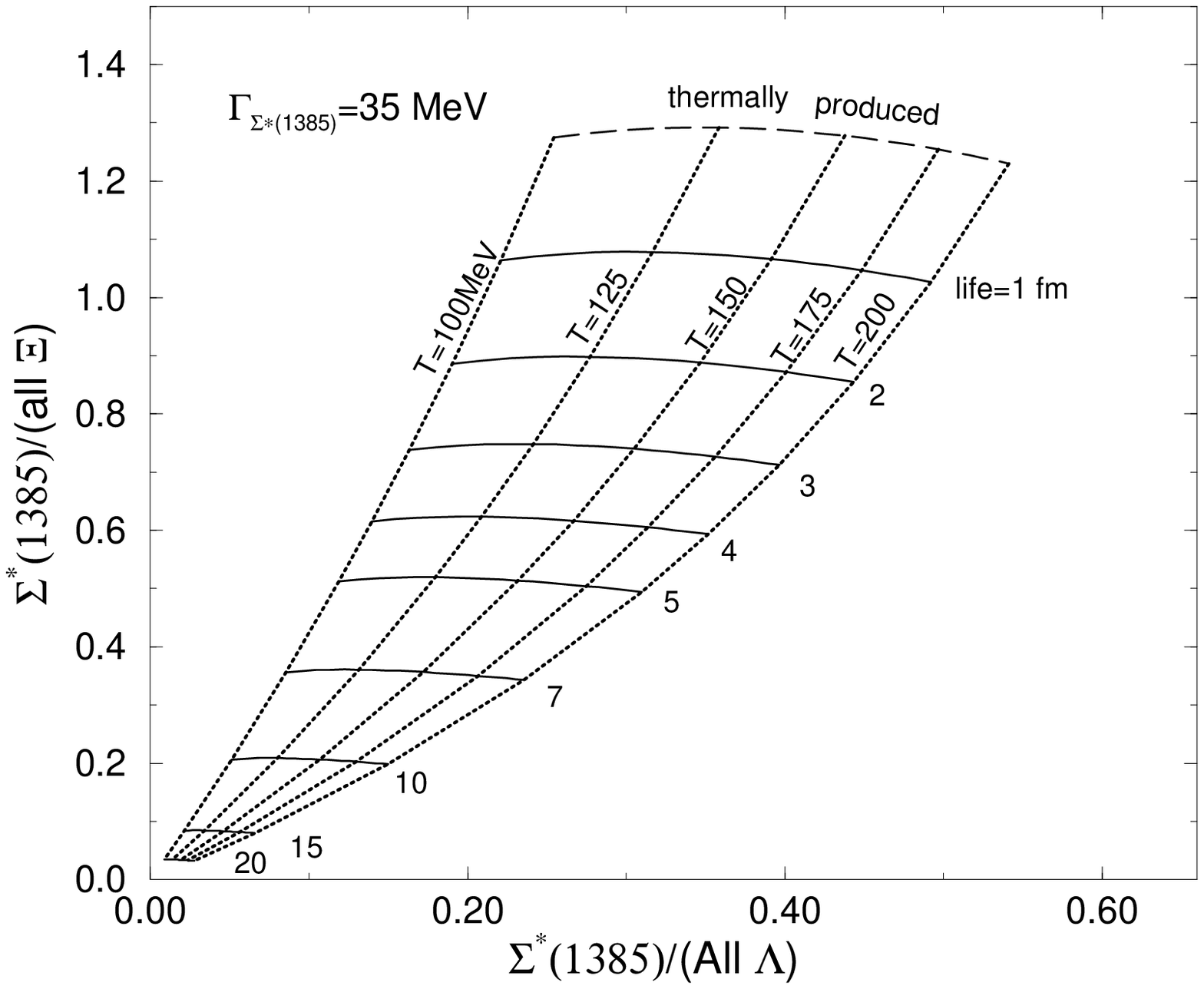 }
\caption{How temperature and fireball lifetime
decouple when two resonances of different masses
and widths are compared.   A point on any of the above diagrams is potentially
sufficient to fix  both of these quantities.
The experimental measurements are discussed in the first section.
See Fig. \ref{prodratios} caption for the meaning of $K,K^*$.}
\label{projdiag}
\end{figure}

Diagrams such as those in Fig. \ref{obsratios} still contain an ambiguity
between temperature and lifetime of the interacting hadron gas phase.
A low observed ratio can either mean a low freeze-out temperature or
a lot of rescattering in a long re-interaction phase.
However, this ambiguity can be resolved by looking at
a selection of resonances, with different masses and lifetimes.
Fig. \ref{projdiag} shows how the initial temperature
and the lifetime of the re-interaction phase decouple when two resonance
ratios are measured simultaneously.
A data point on diagrams such as those in Fig. \ref{projdiag} is enough to
measure both the hadronization temperature and to
distinguish between the sudden freeze-out scenario and a long
re-interaction phase.
\begin{figure}[tb]
\psfig{width=7.3cm,clip=1,figure= 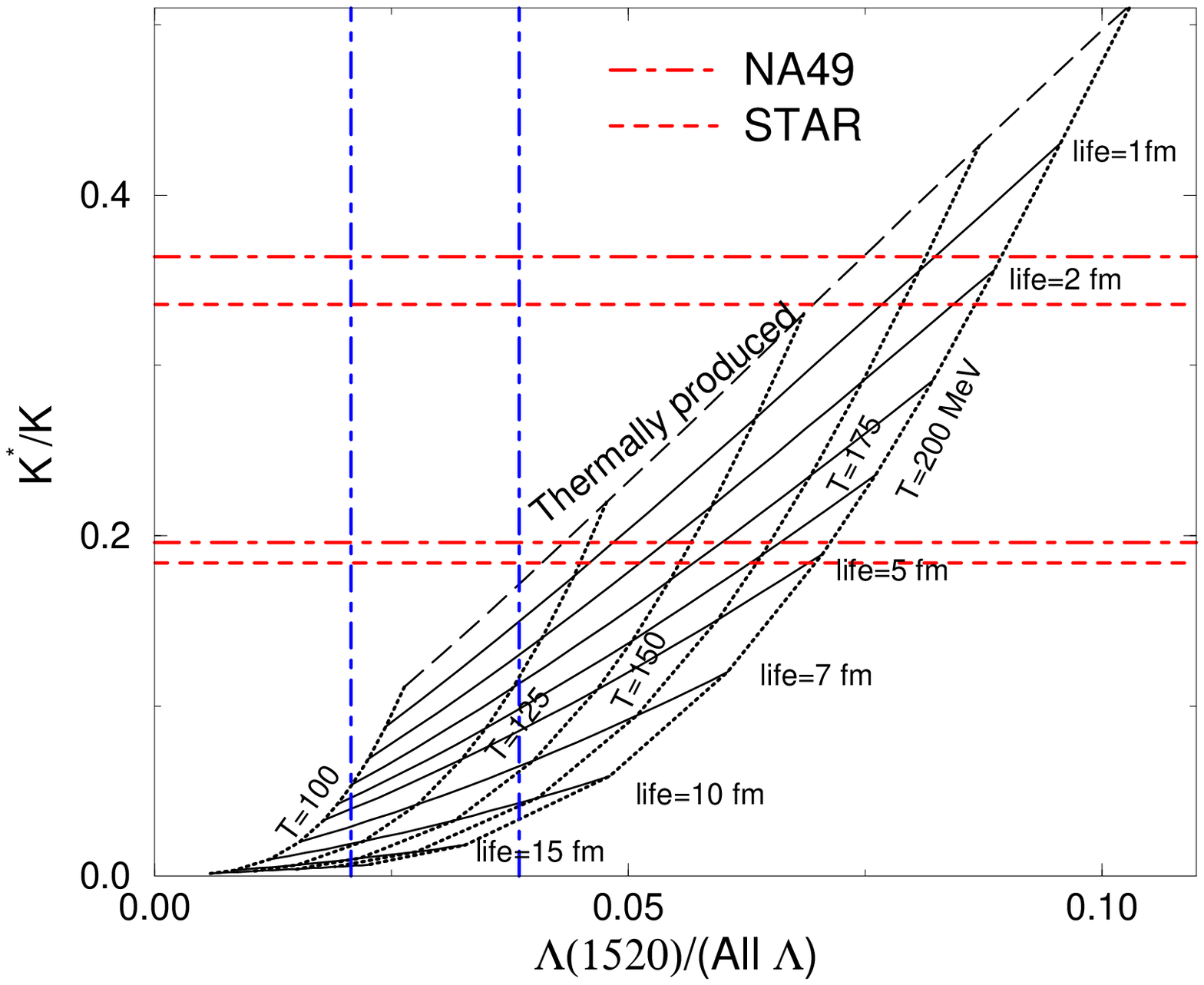}
\psfig{width=7.3cm,clip=1,figure= 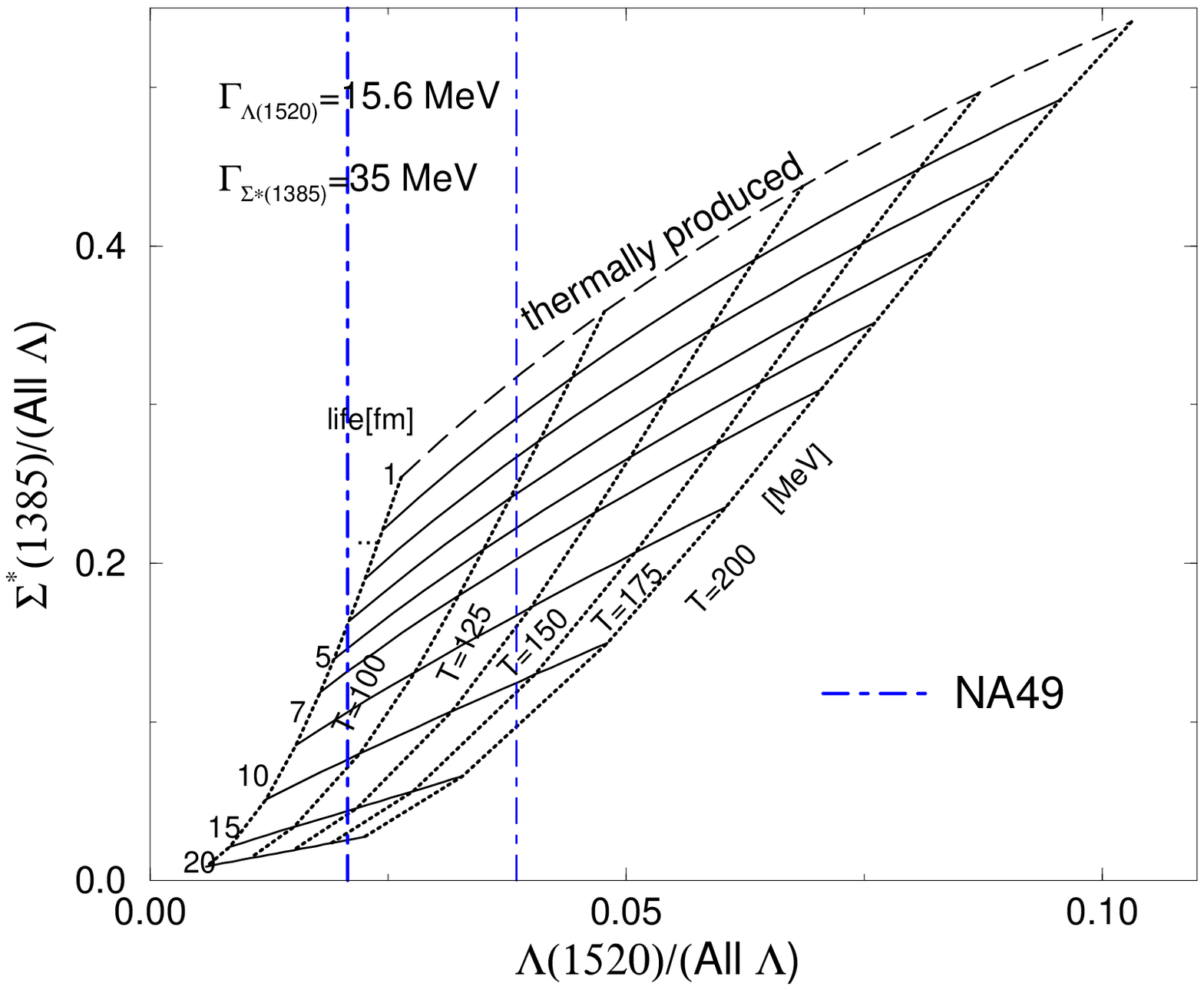 }
\end{figure}
The plots in Fig. \ref{projdiag}
can also be used as consistency checks for the model:
For example, the near independence of $\Sigma^{*}/\Xi$ on temperature
means that the equal temperature lines in the $\Sigma^{*}/\Xi$ vs
 $\Sigma^{*}/\Lambda$ diagram are 
nearly insensitive to the details of the rescattering model.
Moreover the mass differences and lifetimes of the $\Sigma^*$, $K^*$ 
combine in such a way as to make the
$\Sigma^{*}$/(all $\Lambda$) vs $K^{*0}(892)/$(all K-) diagram fold
into a very narrow band.
Any serious shortcoming within our rescattering model would be revealed
if the observed particle ratios stray from this band.
\section{Discussion}
\begin{figure}[tb]
\psfig{width=10cm,clip=1,figure=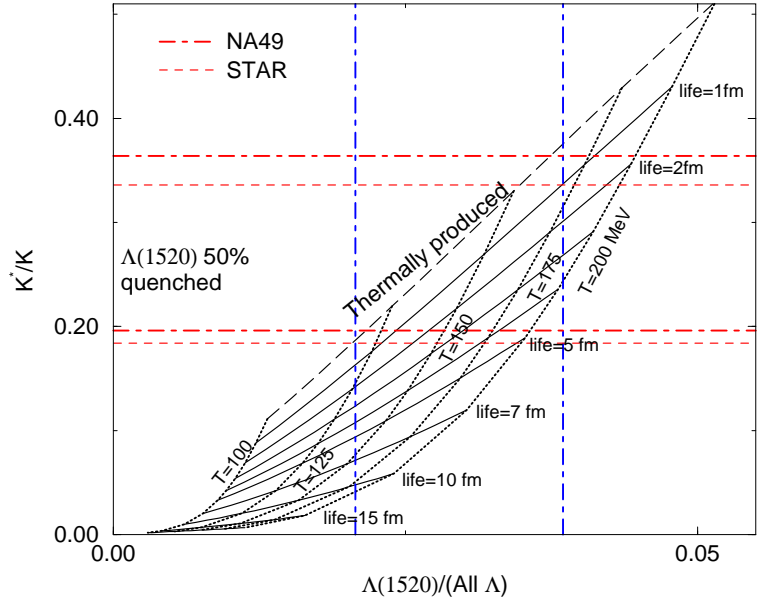 }
\caption{$K^*/K$ vs $\Lambda(1520)/\Lambda$ assuming half 
of the $\Lambda(1520)$ s are suppressed in-medium.   
See Fig. \ref{prodratios} caption for the meaning of $K,K^*$.  }
\label{quenched}
\end{figure}

The first experimental results on short-lived resonances have raised more
questions than answers.
The low $\Lambda(1520)$ multiplicity measured by NA49,
 together with its
non-detection at STAR, can not be understood with a thermal model exclusively.
A $\Lambda(1520)$ suppression factor of at least 2 is
needed to reproduce the measured multiplicity.
However, this suppression does not appear to result from simple
scattering with in-medium particles, described by models
such as UrQMD. Such suppression should manifest
itself much more strongly in case of the $K^{*}$, which has
a much shorter lifetime. 
On the contrary, no comparable depletion
has been found in the case of the $K^{*}$ either at 
SPS or RHIC energies.   Moreover, no broadening of the resonance width, 
which should accompany a strongly interacting medium, has been observed.
If these results are incorporated in the diagrams of Fig. \ref{projdiag}
, the apparent result is sudden freeze-out with an unphysically
low ($\sim 100$ MeV) temperature.
In analysing these results it should be kept in mind 
that the $\Lambda(1520)$ is a very unusual particle:  
 Unlike most other hadronic resonances 
($\Sigma^*,K^*,\Delta,\rho$ ecc.), its high spin is 
due not to valence quark spin configuration
but to the fact that the  $\Lambda(1520)$ s valence quarks are believed to be
in an L=1 state.
The greater space separation of L=1 wave-functions means that the
$\Lambda(1520)$ is especially susceptible to in-medium modifications. 
As Fig. \ref{quenched} shows, a $50 \%$ suppression of the $\Lambda(1520)$
signal at hadronization would mean the data is perfectly compatible
with the sudden freeze-out model described in \cite{Raf00}

Some proposed mechanisms which could lead to a $\Lambda(1520)$ suppression
(Some suggested by Berndt Muller at the meeting \cite{muller}) are:

\begin{itemize}

\item If hadrons form in a quark coalescence process,
the $\Lambda(1520)$ would be suppressed for the same reason that
L=1 (p-shell) electrons are never captured by nuclei \cite{resnick}:
The coalescence probability is dependent on the wave-function overlap
at the formed particle's location 
$\sim |\Psi(x=0)|^2$.
The overlap of two particles goes down for higher L states, and is 0 for odd L=1,3,... states.
However, this effect should also manifest itself in proton-proton
collisions, which serve as basis of the AA-reaction yield. We can thus exclude this effect.

\item Similarly, the larger spatial extent of the $\Lambda(1520)$ means it is more
susceptible to in-medium screening of the strong potential.    This mechanism
invoked for Charmonium suppression 
can also suppress states such as the
$\Lambda(1520)$.

\item In-medium effects can mix
the $\Lambda(1520)$ $|S=3/2,I=1/2 \rangle$ with the $\Sigma^*$ \\$|S=3/2,I=3/2\rangle$ state.
Reactions such as
\begin{equation}
\label{broaden}
\pi + \Lambda(1520) \rightarrow \Sigma^* 
\end{equation}
can deplete the $\Lambda(1520)$ states without a significant effect
on the (much more abundant) $\Sigma^*$ population \cite{torrieri2}.
This is analogous to the phenomenon of quenching, well known in atomic
resonances \cite{urey}:  Resonance radiation from many gases is known to be 
suppressed by in-medium
collisions, which prevent photon emission by converting 
the excitation energy of the resonance into kinetic energy. 

\item
Spin-orbit interactions due to in-medium chromomagnetic fields can also
mix the $\Lambda(1520)$ with the $\Sigma^*$.  

\item Isospin conservation forbids the decay $\Lambda(1520) \rightarrow \Lambda \pi$
but not $\Lambda(1520) \rightarrow \Lambda \sigma$, which is forbidden 
in vacuum since $m_{\Lambda}+m_{\sigma}>m_{\Lambda(1520)}$.
At a temperature high enough for (partial) chiral symmetry restoration to
take place, this may no longer be true, and the 
$\Lambda(1520) \rightarrow \Lambda \sigma$ decay would become possible.
In the same way, partial restoration of chiral symmetry may be
 suppressing the $K^* \rightarrow K \pi$ decay channel.

\item Finally, in a long re-interaction phase there is a possibility that
the $K^*$ will be re-generated.
Processes such as 
\[\ K \pi \rightarrow K^{*} \rightarrow K \pi \]
will lead to an enhancement of observed $K^*$.
Processes which regenerate the $\Lambda(1520)$, such as
\[\ p K \rightarrow \Lambda(1520) \rightarrow p K \] 
are considerably less likely, due to the suppression of higher ($L>0$) partial waves.
\end{itemize}

The last effect can be properly taken into account by a more elaborate rescattering
model, which takes all of the microscopic reaction dynamics into account.
The other mechanisms are rather complex and uncertain, though for the RHIC system, potential 
screening, spin-orbit mixing and $\sigma$
mass change can be explored in a $\mu_B=0$ lattice calculation.
However, these effects also depend on the unusual characteristics
of the $\Lambda(1520)$ resonance. It seems to us that all these
consideration suggest that other resonances
 should be studied to constrain the fireball freeze-out properties.

The non-suppression of the $K^*$ makes it
likely that other hadron resonances can also be detectable.
The diagrams shown in the previous section imply 
that the $\Sigma^*$ should be abundantly produced, and it has many
characteristics which would make it a logical next step in the study
of resonances produced in heavy ion collisions.
Other proposed targets of observation are the $\rho$ and even
the $\Sigma^0 \rightarrow \Lambda \gamma$.

The sudden freeze-out scenario can be thoroughly
tested by the detection of non-strange baryon
resonances ($N^*(1440)$ and $\Delta(1230)$).
The larger widths of these particles, as well as the larger scattering
cross-section in hadronic matter of their decay products, mean that the
very detectability of these particles in terms of an invariant mass
analysis would be strong evidence of a very fast  hadronization process.

\begin{theacknowledgments}
Work supported in part by a grant from the U.S. Department of
Energy,  DE-FG03-95ER40937.  Ch. Markert is supported by 
 the Humboldt foundation.
\end{theacknowledgments}



\end{document}